\documentclass[fleqn,usenatbib,useAMS]{mnras}

\DeclareRobustCommand{\VAN}[3]{#2}
\let\VANthebibliography\thebibliography
\def\thebibliography{\DeclareRobustCommand{\VAN}[3]{##3}\VANthebibliography}

\usepackage{amsmath,amsfonts,amssymb} %for math
\usepackage{bm} %for vectors
\usepackage[table,  dvipsnames]{xcolor} %for author comment colors and that table
\usepackage{hyperref} %to edit citation properties
\usepackage{nicefrac} %for some nice inline fractions
\usepackage{placeins}
\usepackage{microtype}

\usepackage[varg]{txfonts} %aa requirement
\usepackage[T1]{fontenc}

\hypersetup{
	pdftitle = {Spirals, rings, and vortices shaped by shadows in protoplanetary disks: from radiative hydrodynamical simulations to observable signatures},
	pdfauthor = {A. Ziampras, C. P. Dullemond, T. Birnstiel, M. Benisty, R. P. Nelson},
	pdfsubject = {},
	colorlinks = {true},
	linkcolor = [rgb]{0,0.35,0.7},
	citecolor = [rgb]{0,0.35,0.7},
	filecolor = [rgb]{0.61,0,0},
	urlcolor = [rgb]{0.61,0,0},
}

\usepackage{graphicx}  %for images
\newcommand{\tensorGR}[1]{\overline{\bm{{#1}}}}
\newcommand{\DP}[2]{\frac{\partial{#1}}{\partial{#2}}}
\newcommand{\D}[2]{\frac{\text{d}{#1}}{\text{d}{#2}}}

\newcommand{\G}{\text{G}}
\newcommand{\Mstar}{M_\star}
\newcommand{\Lstar}{L_\star}

\newcommand{\Msun}{\mathrm{M}_\odot}
\newcommand{\Lsun}{\mathrm{L}_\odot}

\newcommand{\Rgas}{\mathcal{R}}
\newcommand{\cs}{c_\mathrm{s}}

\newcommand{\OmegaK}{\Omega_\mathrm{K}}

\newcommand{\tauR}{\tau_\mathrm{R}}
\newcommand{\tauP}{\tau_\mathrm{P}}

\newcommand{\taueff}{\tau_\mathrm{eff}}
\newcommand{\kappaR}{\kappa_\mathrm{R}}
\newcommand{\kappaP}{\kappa_\mathrm{P}}

\newcommand{\rhomid}{\rho_\mathrm{mid}}
\newcommand{\sigmaSB}{\sigma_\mathrm{SB}}
\newcommand{\vel}{\bm{u}}

\newcommand{\tcool}{t_\mathrm{cool}}

\newcommand{\Qcool}{Q_\mathrm{cool}}
\newcommand{\Qirr}{Q_\mathrm{irr}}

\newcommand{\Sigmag}{\Sigma_\mathrm{g}}
\newcommand{\Sigmad}{\Sigma_\mathrm{d}}
\newcommand{\velg}{\vel_\mathrm{g}}
\newcommand{\veld}{\vel_\mathrm{d}}
\newcommand{\St}{\mathrm{St}}
\newcommand{\Sc}{\mathrm{Sc}}

\newcommand{\ad}{a_\mathrm{d}}
\newcommand{\brhod}{\bar{\rho}_\mathrm{d}}

\newcommand{\pluto}{\texttt{PLUTO}}

\newcommand{\optool}{\texttt{OpTool}}
\newcommand{\radmc}{\texttt{RADMC-3D}}
\newcommand{\simio}{\texttt{SIMIO-continuum}}
\newcommand{\casa}{\texttt{CASA}}
\title[Shadows, spirals, and rings]{Spirals, rings, and vortices shaped by shadows in protoplanetary disks:\\from radiative hydrodynamical simulations to observable signatures}

\author[A.~Ziampras et al.]{Alexandros~Ziampras$^{1,2,3}$\thanks{E-mail: a.ziampras@lmu.de},
Cornelis~P.~Dullemond$^{4}$,
Tilman~Birnstiel$^{2,5}$,
Myriam~Benisty$^{3,6}$,
Richard~P.~Nelson$^{1}$
\\
$^{1}$Astronomy Unit, School of Physics and Astronomy, Queen Mary University of London, London E1 4NS, UK\\
$^{2}$Ludwig-Maximilians-Universit{\"a}t M{\"u}nchen, Universit{\"a}ts-Sternwarte, Scheinerstr.~1, 81679 M{\"u}nchen, Germany\\
$^{3}$Max Planck Institute for Astronomy, Königstuhl 17, 69117 Heidelberg, Germany\\
$^{4}$Institute for Theoretical Astrophysics, Center for Astronomy (ZAH), Heidelberg University, Albert Ueberle Str.~2, 69120 Heidelberg,
Germany\\
$^{5}$Exzellenzcluster ORIGINS, Boltzmannstr.~2, 85748 Garching, Germany\\
$^{6}$Observatoire de la Côte d'Azur, CNRS, Laboratoire Lagrange, Bd de l'Observatoire, CS 34229, 06304 Nice Cedex 4, France
}

\date{Accepted XXX. Received YYY; in original form ZZZ}
\pubyear{2023}

\begin{document}
\label{firstpage}
\pagerange{\pageref{firstpage}--\pageref{lastpage}}
\maketitle

\begin{abstract}
	Numerous protoplanetary disks exhibit shadows in scattered light observations. These shadows are typically cast by misaligned inner disks and are associated with observable structures in the outer disk such as bright arcs and spirals. Investigating the dynamics of the shadowed outer disk is therefore essential in understanding the formation and evolution of these structures. We carry out twodimensional radiation hydrodynamics simulations that include radiative diffusion and dust--gas dynamics to study the formation of substructure in shadowed disks. We find that spiral arms are launched at shadow edges, permeating the entire disk. The local dissipation of these spirals results in an angular momentum flux, opening multiple gaps and leading to a series of concentric, regularly-spaced rings. We find that ring formation is favored in weakly turbulent disks where dust growth is taking place. These conditions are met for typical class-II disks, in which bright rings should form well within a fraction of their lifetime ($\sim$0.1--0.2\,Myr). For hotter disks gap opening is more efficient, such that the gap edges quickly collapse into vortices that can appear as bright arcs in continuum emission before decaying into rings or merging into massive, long-lived structures. Synthetic observations show that these structures should be observable in scattered light and millimeter continuum emission, providing a new way to probe the presence of substructure in protoplanetary disks. Our results suggest that the formation of rings and gaps is a common process in shadowed disks, and can explain the rich radial substructure observed in several protoplanetary disks.
\end{abstract}


% \bibpunct{(}{)}{;}{a}{}{,}

\section{Introduction}
\label{sec:introduction}

The last decade of observations has revealed that shadows are a common feature in protoplanetary disks around young stars. These shadows, typically theorized to be cast by misaligned inner disks or warps \citep[e.g.,][]{muroarena-etal-2020,bohn-etal-2022,marino-etal-2015}, can span a wide azimuthal extent of the outer disk in scattered light observations when the inner disk is only tilted by a few degrees, and are associated with a variety of observed structures such as arcs and spirals. Such examples include HD~139614 \citep{muroarena-etal-2020}, HD~143006 \citep{benisty-etal-2018}, and Wray~15-788 \citep{bohn-etal-2019}. In the case of highly inclined inner disks, the shadow can be more localized and appear as a dark lane, blocking incident stellar radiation in a narrow azimuthal range. Examples of narrow shadows include HD~100453 \citep{benisty-etal-2017}, HD~135344B \citep{stolker-etal-2017}, DoAr~44 \citep{avenhaus-etal-2018,casassus-etal-2018}, and RXJ1604.3-2130A \citep{pinilla-etal-2018}.

The presence of one or more shadows or warps in a protoplanetary disk has been shown to significantly influence both its thermal and chemical structure \citep{young-etal-2021} as well as its dynamical evolution \citep[e.g.,][]{juhasz-facchini-2017,kimmig-dullemond-2024}. Gas in the shadowed region can cool quite efficiently, especially so in the optically thin outer disk \citep[e.g.,][]{casassus-etal-2019,nealon-etal-2020}. The resulting temperature contrast between the shadowed and illuminated regions translates to a radial force imbalance, as pressure support is weaker inside the shadow, which manifests as spirals in the disk \citep{montesinos-etal-2016,montesinos-cuello-2018,zhang-zhu-2024}. These spirals provide an additional interpretation to spiral structures besides embedded planets \citep[e.g.,][]{zhu-etal-2015} and turbulence due to the gravitational instability \citep[e.g.,][]{gammie-2001,bethune-latter-2021}.

At the same time, such vigorous spiral activity would provide an efficient mechanism to redistribute angular momentum within the disk. The nonlinear damping of spiral arms can drive a local angular momentum flux into the disk due to them steepening into shocks \citep[e.g.,][]{rafikov-2002} at the interface between the shadowed and illuminated regions, or due to radiative cooling \citep{miranda-rafikov-2020a,zhang-zhu-2020,ziampras-etal-2023a}. This process essentially acts to remove material from the location where damping occurred, ultimately leading to the formation of a gap if the angular momentum flux is both sustained long enough at a given radial location and strong enough to overcome the gap-refilling effects of turbulence and local pressure gradients \citep{crida-etal-2006}. Due to the long timescales involved in the gap-opening process, however, the possibility of gap opening due to the spiral arms excited in shadowed disks has not been explored in detail \citep[but see][whose work was published during the preparation of this manuscript]{su-bai-2024}.

In this work we aim to investigate the formation of radial structure in the form of rings and gaps due to the spiral wakes excited in a protoplanetary disk shadowed by a misaligned inner disk. We approach this problem with radiation hydrodynamics simulations that include a self-consistent treatment of cooling and radiative diffusion, coupled to a dust component that accounts for the dynamics of both submicron and millimeter grains. In this way, we can study both the thermal and dynamical evolution of the gas structure in the presence of a shadow, as well as the observability of any resulting substructure in scattered light or millimeter continuum observations.

In Sect.~\ref{sec:physics-numerics} we introduce our physical and numerical framework and describe our model of a shadowed protoplanetary disk. We analyze our results from a suite of hydrodynamical simulations in Sects.~\ref{sec:fiducial}~\&~\ref{sec:parameter-study}, and present a set of synthetic observations in Sect.~\ref{sec:observations}. We discuss our findings in Sect.~\ref{sec:discussion}, and finally conclude in Sect.~\ref{sec:summary}.

\section{Physics and numerics}
\label{sec:physics-numerics}

In this section we introduce our physical framework and describe the approach we take to model a shadowed protoplanetary disk. We also outline our numerical setup and list the parameters we use in our simulations.

\subsection{Radiation hydrodynamics}
\label{sub:rad-hydro}

We solve the Navier--Stokes equations \citep[e.g.,][]{tassoul-1978} in cylindrical coordinates $\{R,\phi\}$ for a vertically-integrated distribution of perfect gas with adiabatic index $\gamma=7/5$ and mean molecular weight $\mu=2.35$ orbiting around a star with mass $\Mstar$ and luminosity $\Lstar$. We also include a distribution of dust grains with bulk density $\brhod$ and size $\ad$, modeled as a pressureless fluid. Defining the gas and dust surface density $\Sigmag$ and $\Sigmad$, the gas and dust velocities $\velg$ and $\veld$, and the vertically integrated gas pressure $P$, the equations read
\begin{subequations}
	\label{eq:navier-stokes}
	\begin{align}
		\label{eq:navier-stokes-1}
		\D{\Sigmag}{t} = -\Sigmag\nabla\cdot\velg,
	\end{align}
	\begin{align}
		\label{eq:dust-evolution-1}
		\D{\Sigmad}{t} = -\Sigmad\nabla\cdot\veld -\nabla\cdot\bm{j},\quad\bm{j}=-\Sigma\frac{\nu}{\Sc}\frac{1+4\St^2}{(1+\St^2)^2}\nabla\left(\frac{\Sigmad}{\Sigma}\right)
	\end{align}
	\begin{align}
	\label{eq:navier-stokes-2}
		\Sigmag\D{\velg}{t} =-\nabla P -\Sigmag\nabla\Phi_\star +\nabla\cdot\bm{\tensorGR{\sigma}} -\Sigmad\frac{\velg-\veld}{\St}\OmegaK,
	\end{align}
	\begin{align}
		\label{eq:dust-evolution-2}
		\Sigmad\D{\veld}{t} = - \Sigmad\nabla\Phi_\star - \Sigmad\frac{\veld-\velg}{\St}\OmegaK,
	\end{align}
	\begin{align}
	\label{eq:navier-stokes-3}
		\D{e}{t}=-\gamma e\nabla\cdot\velg+Q_\mathrm{visc}+Q_\mathrm{irr} + Q_\mathrm{cool} + Q_\mathrm{rad}.
	\end{align}
\end{subequations}
Here, $\Sigma=\Sigmag+\Sigmad$ is the total surface density, $e=P/(\gamma-1)$ is the internal energy density, $\Phi_\star=-\G\Mstar/R$ is the gravitational potential of the central star of mass $\Mstar$ at distance $R$, $\tensorGR{\sigma}$ is the viscous stress tensor, and $\OmegaK=\sqrt{\G\Mstar/R^3}$ is the Keplerian angular velocity, with $\G$ being the gravitational constant. We can then define the gas temperature $T=\mu \cs^2/\Rgas$ with $\cs=\sqrt{P/\Sigmag}$ the isothermal sound speed, which relates to the pressure scale height $H=\cs/\OmegaK$. The disk aspect ratio is then $h=H/R$.

In Eq.~\eqref{eq:dust-evolution-1}, $\bm{j}$ represents the dust diffusion flux \citep{morfill-voelk-1984,youdin-lithwick-2007} for a kinematic viscosity $\nu$ and assuming that the Schmidt number is $\Sc=1$. Finally, $\St$ denotes the Stokes number or dimensionless stopping time of the dust grains, which are subject to Epstein drag \citep[e.g.,][]{armitage-2009} with
\begin{equation}
	\label{eq:stokes-number}
	\St = \frac{\pi}{2}\frac{\ad\brhod}{\Sigmag}.
\end{equation}

The terms $Q_\mathrm{visc}$, $Q_\mathrm{irr}$, $Q_\mathrm{cool}$, and $Q_\mathrm{rad}$ represent viscous heating, irradiation heating, surface cooling, and in-plane radiative diffusion, respectively:
\begin{subequations}
	\label{eq:source-terms}
	\begin{align}
		\label{eq:source-terms-1}
		Q_\mathrm{visc} = \frac{1}{2\nu\Sigmag}\mathrm{Tr}(\tensorGR{\sigma}^2) \approx \frac{9}{4}\nu\Sigmag\OmegaK^2,\quad \nu=\alpha\sqrt{\gamma}\cs H,
	\end{align}
	\begin{align}
		\label{eq:source-terms-2}
		Q_\mathrm{irr} = 2\frac{\Lstar}{4\pi R^2} (1-\epsilon)\frac{\theta}{\taueff}, \quad \theta = R\D{h}{R}\approx \frac{2h}{7},
	\end{align}
	\begin{align}
		\label{eq:source-terms-3}
		Q_\mathrm{cool} = -2\frac{\sigmaSB T^4}{\taueff}, \quad \taueff = \frac{3\tauR}{8} + \frac{\sqrt{3}}{4} + \frac{1}{4\tauP}, \quad \tau_\text{K,P} = \frac{\kappa_\text{K,P}\Sigmag}{2},
	\end{align}
	\begin{align}
		\label{eq:source-terms-4}
		Q_\mathrm{rad} = \sqrt{2\pi}H\nabla\cdot \left(\lambda\frac{4\sigmaSB}{\kappaR\rhomid}\nabla T^4\right), \quad \rhomid = \frac{1}{\sqrt{2\pi}}\frac{\Sigmag}{H}.
	\end{align}
\end{subequations}
In the above, we follow the $\alpha$-viscosity prescription of \citet{shakura-sunyaev-1973}, the irradiation model of \citet{menou-goodman-2004} for a star with luminosity $\Lstar$ assuming a disk albedo $\epsilon=1/2$, an effective optical depth $\taueff$ that depends on the Rosseland and Planck mean opacities $\kappaR$ and $\kappaP$ according to \citet{hubeny-1990}, and the flux-limited diffusion (FLD) closure of \citet{levermore-pomraning-1981} for the radiative diffusion term. We use the prescription by \citet{kley-1989} for the flux limiter $\lambda$. Finally, $\sigmaSB$ denotes the Stefan--Boltzmann constant. For more details on the individual radiative terms, we refer the reader to \citet{ziampras-etal-2023a}.

\subsection{Dust model, parameters, and initial conditions}
\label{sub:parameters}

We consider a solar-type star with $\Mstar=1\,\Msun$ and $\Lstar=1\,\Lsun$ and a disk with surface density profiles $\Sigmag=1000\,R_\text{au}^{-1}\,\text{g}/\text{cm}^2$ and $\Sigmad=0.01\Sigmag$ for the gas and dust, respectively. This translates to a disk mass of $\sim0.07\,\Msun$ assuming a disk that extends between 0.1--100\,au. For the dust grains we assume a two-population model similar to \citet{ziampras-etal-2024b}, where the total dust mass is distributed between ``small'' and ``big'' grains with $\ad^\text{small}=0.1\,\mu\text{m}$ and $\ad^\text{big}=1\,\text{mm}$. For convenience, we define a coagulation fraction $X$ and write
\begin{equation}
	\label{eq:coagulation-fraction}
	X = \Sigmad^\text{big}/\Sigmad,\qquad \Sigmad = \Sigmad^\text{small}+\Sigmad^\text{big}.
\end{equation}
In this way, $X=0$ and $X=1$ correspond to disks containing only small or big grains, respectively.
The small grains are assumed to be perfectly coupled to the gas and at any given time their surface density is given by $\Sigmad^\text{small} = (1-X_0)\varepsilon_0\Sigmag$, where $X_0$ and $\varepsilon_0=0.01$ are the coagulation fraction and dust-to-gas ratio at $t=0$. The big grains are initialized as $\Sigmad^\text{big} = X_0\varepsilon_0\Sigmag$, and evolved according to Eqs.~\eqref{eq:dust-evolution-1}~\&~\eqref{eq:dust-evolution-2}.

The Rosseland and Planck mean opacity of each grain population is given by
\begin{align}
	\label{eq:opacities}
	&\kappaR^\text{small} \approx 0.27~T_\text{K}^{1.6}\,\text{cm}^2/\text{g},\quad \kappaP^\text{small} \approx 0.41~T_\text{K}^{1.6}\,\text{cm}^2/\text{g},\\
	&\kappaR^\text{big} \approx \kappaP^\text{big}\approx 4.5\,\text{cm}^2/\text{g},
\end{align}
following the calculations by \citet{ziampras-etal-2024b} with the code \optool{} \citep{dominik-etal-2021}, using the Distribution of Hollow Spheres approach \citep[DHS,][]{min-etal-2005}. The dust grains are composed of 87\% amorphous pyroxenes and 13\% amorphous carbon \citep{zubko-etal-1996} with a porosity of 25\%, resulting in $\brhod=2.08\,\text{g}/\text{cm}^3$. The total opacity per gram of gas to be used in Eqs.~\eqref{eq:source-terms-2}--\eqref{eq:source-terms-4} is then
\begin{equation}
	\label{eq:total-opacity}
	\kappa_\text{R,P} = \frac{\Sigmad^\text{small}\kappa_\text{R,P}^\text{small} + \Sigmad^\text{big}\kappa_\text{R,P}^\text{big}}{\Sigmag} = \varepsilon \left[(1-X)\kappa_\text{R,P}^\text{small} + X\kappa_\text{R,P}^\text{big}\right].
\end{equation}
As shown in \citet{ziampras-etal-2024b}, this approach works well even for marginally optically thick regions where the Rosseland mean opacity becomes important.

For a small $\alpha=10^{-5}$ and our disk model, Eq.~\eqref{eq:navier-stokes-3} yields in steady state $h\approx0.019\,R_\text{au}^{2/7}$, or $T\approx90\,R_\text{au}^{-3/7}\,\text{K}$ \citep[see e.g.,][]{chiang-goldreich-1997} regardless of our choice of $X$ due to heating from viscous dissipation being negligible. Translating the above to a reference radius $R_0=30\,$au, our initial conditions are then
\begin{subequations}
	\label{eq:initial-conditions}
	\begin{align}
		&\Sigma_\text{g,0} = 33.3\,\text{g}/\text{cm}^2\,\left(\frac{R}{R_0}\right)^{-1},\quad \Sigma_\text{d,0}^\text{big} = X_0\varepsilon_0\Sigma_\text{g,0},\\
		&h_0 = 0.05\,\left(\frac{R}{R_0}\right)^{2/7}\Rightarrow T_0 = 20.9\,\text{K}\,\left(\frac{R}{R_0}\right)^{-3/7}.
	\end{align}
\end{subequations}

In our models we choose a fiducial value of $X=0.9$ (i.e., 90\% of the dust mass is in big grains) but also consider $X=\{0.1,0.99\}$ to investigate the effect of dust growth on both radiative processes and the observability of any resulting substructure. We also explore the effects of diffusion by varying $\alpha=\{10^{-5}, 10^{-4}, 10^{-3}\}$, with $10^{-5}$ being our fiducial value.

Finally, we consider three different disk aspect ratios by varying $h(R_0)=\{0.05,0.07,0.1\}$, with $0.05$ being our fiducial value. To achieve this we simply adjust the stellar luminosity $\Lstar$ in Eq.~\eqref{eq:source-terms-2} to $\{1, 10.6, 128.7\}\,\Lsun$ for $h=\{0.05,0.07,0.1\}$, respectively. While this is of course unrealistic, it allows us to investigate both the effect of a higher sound speed on the formation of spirals and the importance of radiative cooling on any resulting substructure, as the cooling timescale (loosely) scales with $\tcool\propto e/|\Qcool| \propto T^{-3} \propto h^{-6}$.

A list of all models carried out in this work can be found in Table~\ref{table:models}.

\begin{table}
	\begin{center}
	\caption{List of hydrodynamical models in this study. Values for $X$ refer to $t=0$, and values for $h_0$ refer to $t=0$ and $R=R_0=30$\,au.}
	\label{table:models}
	\begin{tabular}{c|c|c|c}
		$X$  & $\alpha$ & $h_0$ & comment \\
		\hline
		0.9  & $10^{-5}$ & 0.05 & fiducial model\\
		0.1  & $10^{-5}$ & 0.05 & fewer mm grains (10\% by mass)\\
		0.99 & $10^{-5}$ & 0.05 & more mm grains (99\% by mass)\\
		0.9  & $10^{-4}$ & 0.05 & moderately viscous \\
		0.9  & $10^{-3}$ & 0.05 & highly viscous \\
		0.9  & $10^{-5}$ & 0.07 & moderately hotter disk ($\Lstar=10.6\,\Lsun$)\\
		0.9  & $10^{-5}$ & 0.1  & much hotter disk ($\Lstar=128.7\,\Lsun$)\\
	\end{tabular}
	\end{center}
\end{table}

\subsection{Shadows due to a misaligned inner disk}
\label{sub:shadow-model}

Similar to \citet{montesinos-etal-2016} and \citet{cuello-etal-2019}, we consider a small, misaligned inner disk casting a shadow on the ``main'' disk that we actually evolve in our simulations. We assume that the inner disk extends between $R\in[0.2,4]$\,au and is tilted by $30^\circ$ with respect to the $xy$ plane about the $x$ axis, while the main disk extends beyond 5\,au. The two disks otherwise share the same physical properties (see Sec.~\ref{sub:parameters}), and we apply an exponential taper in the radial direction for both disks for a smooth transition between the two. A detailed description of our two-disk model is provided in Appendix~\ref{apdx:two-disks}.

\begin{figure}
	\includegraphics[width=\columnwidth]{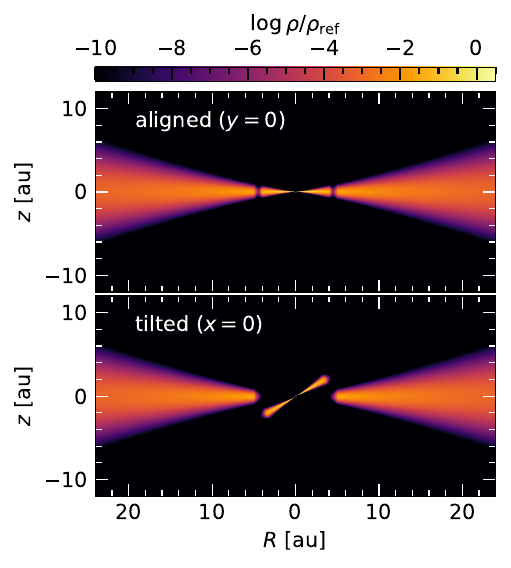}
	\caption{Schematic of the disk configuration with slices along the $x$ (top) and $y$ axes (bottom). The inner disk is tilted about the $x$ axis by $30^\circ$ with respect to the $xy$ plane and casts a shadow on the main disk.}
	\label{fig:setup}
\end{figure}

We construct this model in \texttt{Python} using a 3D spherical grid $\{r,\theta,\phi\}$ logarithmically spaced in $r$ with $r\in[0.1,300]$\,au, $\theta\in[0,\pi]$, $\phi\in[-\pi,\pi]$ and $N_r\times N_\theta\times N_\phi = 512\times768\times512$ cells. A plot of the configuration is shown in Fig.~\ref{fig:setup}.

To measure the width and depth of the shadow cast by the inner disk we then trace rays from the central star along the radial direction and measure the irradiation heating term $\mathcal{S}$
\begin{equation}
	\label{eq:stellar-heating}
	\mathcal{S} \approx \frac{\Lstar e^{-\tau_\text{irr}(r)} }{4\pi r^2} \left(1-e^{-\kappaP^\text{small}\rho_\text{d}^\text{small}\Delta r}\right), \quad \tau_\text{irr}(r) = \int\limits_{0.1\,\text{au}}^{r}\kappaP^\text{small}\rho_\text{d}^\text{small}\,dr.
\end{equation}
Here, $\Delta r$ is the radial cell width, and $\rho_\text{d}^\text{small}$ is the volume density in small dust, assumed to be perfectly coupled to the gas and therefore given by $\rho_\text{d}^\text{small}(r,\theta)=\varepsilon (1-X)\rho_\text{g}$, where $\rho_\text{g}$ is the gas density (see Eq.~\eqref{eq:dust-density}). We set $X=0.9$ and $\varepsilon=0.01$.

\begin{figure}
	\includegraphics[width=\columnwidth]{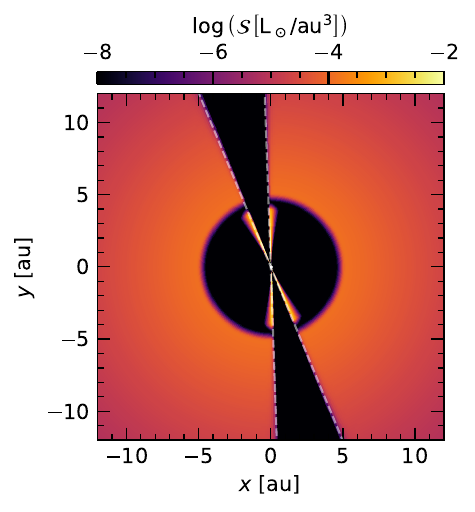}
	\caption{Heatmap of the irradiation heating term $\mathcal{S}$ through Eq.~\eqref{eq:stellar-heating} along a surface tilted by $7^\circ$ with respect to the $xy$ plane, tracing the irradiation surface of the main disk.
	The shadow cast by the inner disk is clearly visible as a dark wedge. Dashed lines mark the shadow edges, fit by eye.}
	\label{fig:shadow}
\end{figure}

\begin{figure}
	\includegraphics[width=\columnwidth]{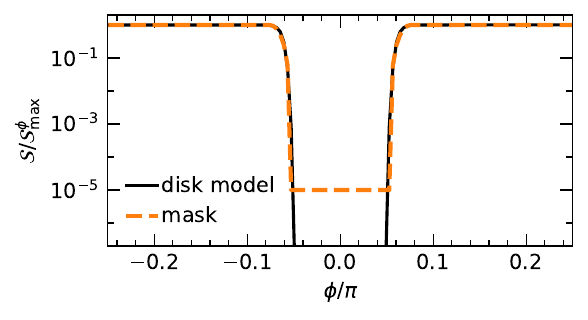}
	\caption{Azimuthal profile of the irradiation heating term $\mathcal{S}$ at $z=0$ and $R=10$\,au. The mask $f_\mathrm{sh}$ used in our disk models, defined in Eq.~\eqref{eq:mask}, is shown in orange.}
	\label{fig:mask}
\end{figure}

In Fig.~\ref{fig:shadow} we show a heatmap of $\mathcal{S}$ along a surface tilted by $4\,h_{5\,\text{au}}\approx 7^\circ$ with respect to the $xy$ plane about the $x$ axis, roughly tracing the irradiation surface of the main disk in order to showcase the geometry of the shadow.
From this figure it becomes clear that inside the shadow the irradiation heating is reduced to practically zero, while outside the shadow the heating is unaffected. We can then define a mask that we apply to the irradiation heating term $\Qirr$ in Eq.~\eqref{eq:source-terms-3} to simulate the shadow cast by the inner disk. This mask is compared to the heating term $\mathcal{S}$ at the midplane in Fig.~\ref{fig:mask} and is given by
\begin{equation}
	\label{eq:mask}
		f_\mathrm{sh} = \begin{cases}
			d_\mathrm{sh}, & |x| \leq 0.7 \\
			d_\mathrm{sh} + (1-d_\mathrm{sh}) \sin\left(\frac{|\phi-\phi_\mathrm{e}|}{0.3\,w_\mathrm{sh}}\frac{\pi}{2}\right)^2, & 0.7 < |x| < 1 \\
			1, & |x| \geq 1 \\
		\end{cases}
\end{equation}
with $x=(\phi-\phi_0)/w_\mathrm{sh}$ and $\phi_\mathrm{e} = \phi_0 + \mathrm{sgn}(x)/0.7$. The shadow depth and half-width are set to $d_\mathrm{sh}=10^{-5}$ and $w_\mathrm{sh}=0.24$, respectively. We model two diametrically opposed shadows at $\phi_0=0$ and $\phi_0=\pi$. The mask is then applied to the irradiation heating term $\Qirr$ in Eq.~\eqref{eq:source-terms-3} as
\begin{equation}
	\label{eq:shadowed-heating}
	\Qirr \to \Qirr f_\mathrm{sh}.
\end{equation}

We note that we are not interested in the absolute value of $\mathcal{S}$ but rather its azimuthal dependence, which we then use to mask the main disk in our simulations. Since the disk model has already been established in Sec.~\ref{sub:parameters} (i.e., we do not actually calculate the disk thermal structure through ray-tracing), our results are only weakly sensitive to the opacity model. For this reason, we use the same mask for all models instead of recomputing it for different $X$. To isolate the effects of different $h$ in the dynamics of the disk, we also use the same mask for all $h$ even though the shadow would realistically be wider for a geometrically thicker disk.

\subsection{Numerical setup}
\label{sub:numerics}

We solve the 2D, vertically integrated set of Eqs.~\eqref{eq:navier-stokes} using the numerical hydrodynamics code \pluto{} \citep{mignone-etal-2007}, combined with the dust module described in \citet{ziampras-etal-2024b}, the FLD module in \citet{ziampras-etal-2020a}, and a dust diffusion module following \citet{weber-etal-2019}. To relax timestep limitations we use the FARGO algorithm \citep{masset-2000}, implemented in \pluto{} by \citet{mignone-etal-2012}. We use a second-order time- and spatially-accurate scheme with the HLLC Riemann solver \citep{toro-etal-1994} and the flux limiter of \citet{vanleer-1974}. Viscosity is implemented using the super-time-stepping method of \citet{alexiades-etal-1996}.

Our grid extends radially with a logarithmic spacing between $R\in[6,75]\,\text{au} = [0.2,2.5]\,R_0$ with a reference radius of $R_0=30$\,au and covers the full azimuthal extent $\phi\in[0,2\pi]$ with $N_r\times N_\phi = 576\times1536$ cells. This grid resolution translates to approximately 12 cells per gas scale height in either direction at $R=R_0$ for $h_0=0.05$. Our disk is initialized according to Eq.~\eqref{eq:initial-conditions}, with $u_R=0$ for both gas and dust and
\begin{equation}
	\label{eq:vphi}
	u_{\phi,\text{g}}=\OmegaK R \sqrt{1+\D{\log P}{\log R} h^2}, \quad u_{\phi,\text{d}}=\OmegaK R,
\end{equation}
to correct for pressure support.

We impose wave-damping boundary conditions at both radial boundaries on all quantities following \citet{devalborro-etal-2006} with a damping timescale of 0.1 boundary orbits. At the boundary walls all quantities are fixed to their initial values. We then evolve the system for 1000 orbits at $R_0$ ($=1000\,P_0$), which corresponds to approximately 164\,kyr ($P_0\approx 164$\,yr).

\section{Fiducial model}
\label{sec:fiducial}

In this section we present the results of our fiducial numerical simulation with $X=0.9$, $\alpha=10^{-5}$, and $h(R_0)=0.05$. This set of parameters comprises a reasonably realistic disk model with a significant fraction of the dust mass in big grains after substantial dust growth has taken place \citep[see e.g.,][]{birnstiel-2023}, a small viscosity motivated by the lack of magnetically-driven turbulence \citep[e.g.,][]{bai-stone-2013}, and a temperature profile that corresponds to a passively irradiated disk \citep{chiang-goldreich-1997}.

\subsection{Disk structure}
\label{sub:disk-structure}

\begin{figure*}
	\includegraphics[width=\textwidth]{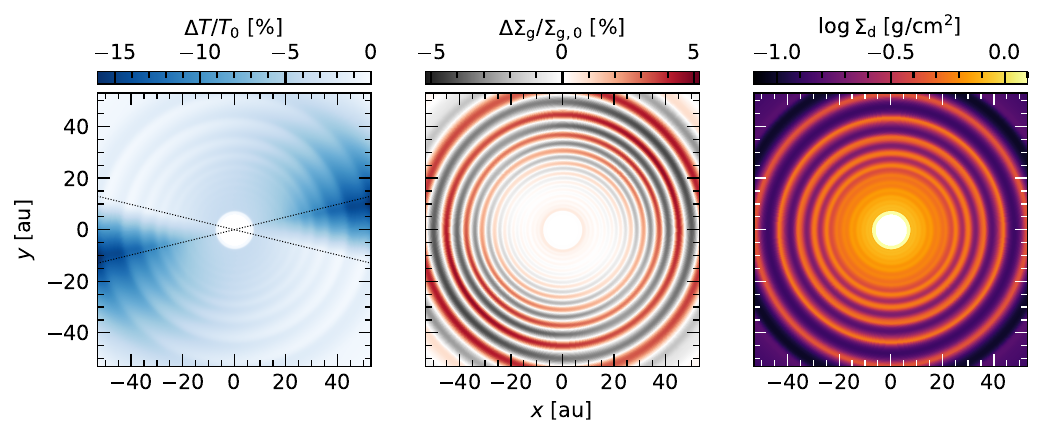}
	\caption{The perturbed gas temperature (left) and gas surface density (middle) as well as the dust surface density (right) after $1000\,P_0$. Two dark wedges are visible in the temperature structure, corresponding to the attenuated heating due to the ``shadows'' cast by an inner disk. A series of concentric rings is visible in all quantities, forming clearly visible features in the dust surface density. The shadow edges are marked with dashed lines in the left panel. The disk rotates counterclockwise.}
	\label{fig:fiducial}
\end{figure*}
We start by showing heatmaps of the perturbed temperature and surface density of the disk after $1000\,P_0$ of Fig.~\ref{fig:fiducial}. The edges of the two shadows are marked with dotted lines in the perturbed temperature (left panel), with the gas cooling rapidly once it enters a shadow and then heating back up to the irradiation temperature upon exiting. The temperature troughs are azimuthally offset with respect to the center of each shadow as the gas responds to heating and cooling over a finite timescale.

\begin{figure}
	\includegraphics[width=\columnwidth]{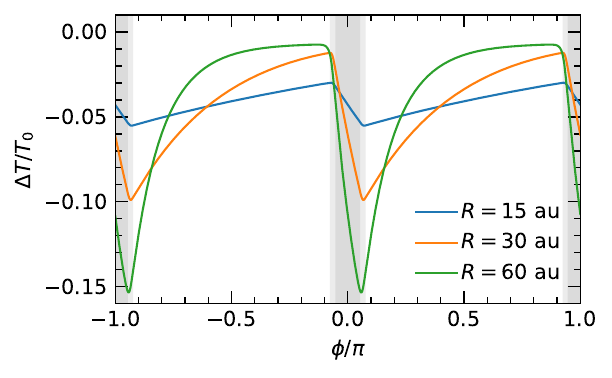}
	\caption{Azimuthal profile of the perturbed gas temperature at different radii at $t=1000\,P_0$. The temperature perturbation is not symmetric about the shadow exits, and becomes smaller in the slowly-cooling inner disk. Shaded regions mark the two shadows. Gas moves to the right.}
	\label{fig:fiducial-T}
\end{figure}

The azimuthal profile of temperature at various radii is shown in Fig.~\ref{fig:fiducial-T}, further revealing that the temperature perturbation is not symmetric about the exit of the shadow. This happens due to radiative cooling being proportional to $T^4$, and results in the gas cooling more rapidly than it heats up. We can also see that the gas does not quite reach the irradiation temperature before entering the next shadow, leading to a net cooling of the disk by a few percent compared to the initial state. This effect is more pronounced in the inner disk, where the cooling timescale is longer due to the higher gas density and opacity. The slower cooling also results in a smaller temperature contrast between the shadowed and illuminated regions. Specifically, as the cooling timescale increases, the azimuthal temperature perturbation becomes smaller and approaches a value of $\approx 4\%$, reflecting the fact that the disk is irradiated less on average due to the pair of shadows.

Interestingly, while expecting a spiral structure in the gas surface density (middle panel of Fig.~\ref{fig:fiducial}) as the result of the radial force imbalance due to the shadow, we instead find a series of concentric rings. Even though the contrast of these rings is of the order of 5\% in the gas surface density, the local modulation of the pressure gradient is enough to significantly enhance the dust-to-gas ratio at the peak of each ring over time, as shown in the dust surface density (right panel).

We speculate that the formation of rings is inherently linked to the launching of spiral waves at the shadow edges, due to the lower temperature in the shadowed region resulting in a radial force imbalance (see Eq.~\eqref{eq:vphi}). If these spiral waves can dissipate into the disk, the resulting angular momentum flux (AMF) can drive the formation of gaps in the gas surface density. This in turn leads to a smaller radial pressure gradient between two gaps, potentially trapping---or at least slowing down---dust grains. The result is a series of rings in dust continuum emission. 

\begin{figure*}
	\includegraphics[width=\textwidth]{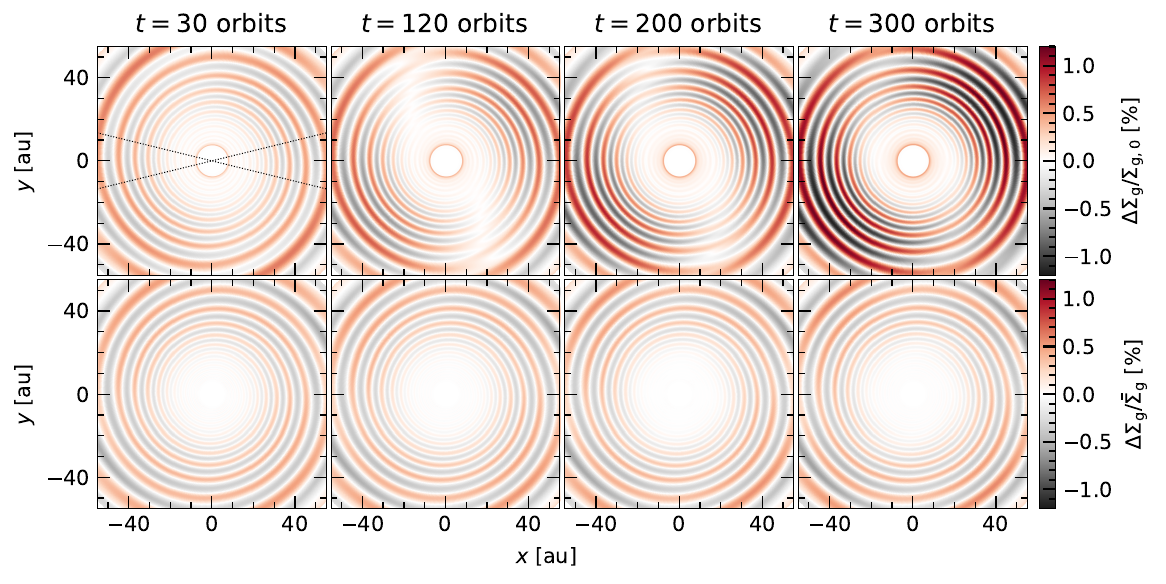}
	\caption{The perturbed gas surface density with respect to the initial state $\Sigma_\text{g,0}$ (top panels) and the azimuthally averaged state $\bar{\Sigma}_\text{g}$ (bottom panels) at different snapshots. The disk initially shows spiral-like structures, but transitions to the rings visible in Fig.~\ref{fig:fiducial} after about $120\,P_0$. The undulations seem to be more pronounced at the exits of the shadows and increase in amplitude over time.}
	\label{fig:fiducial-spiral-ring}
\end{figure*}

Figure~\ref{fig:fiducial-spiral-ring} shows the perturbed gas surface density with respect to the initial state $\Sigma_\text{g,0}$ (top panels) and the azimuthally averaged state $\bar{\Sigma}_\text{g}$ (bottom panels) at different snapshots. At early times, an $m=2$ spiral pattern can be seen in both top and bottom panels, owing to the radial force imbalance in the shadow perturbing the initially axisymmetric disk. This spiral pattern persists in the bottom panels, indicating that spirals are continuously generated throughout the disk. In the top panels, however, we observe a transition from spiral- to ring-like structures, with the rings becoming more pronounced over time. The transition seems to take place after about $120\,P_0$ (second column), where we can see what looks like spirals breaking up before merging into rings.

This behavior supports our hypothesis that the constant generation of spiral waves due to the shadows drives an AMF into the disk once the waves dissipate, ultimately leading to the formation of rings. We investigate further by computing the AMF in the disk, defined as \citep[e.g.,][]{miranda-rafikov-2020a}
\begin{equation}
	\label{eq:amf}
	F_J = \oint\limits_\phi \delta F_J \text{d}\phi, \quad \delta F_J(R,\phi) = \bar{\Sigma} R^2 (u_R-\bar{u}_R)\,(u_\phi - \bar{u}_\phi),
\end{equation}
\begin{figure}
	\includegraphics[width=\columnwidth]{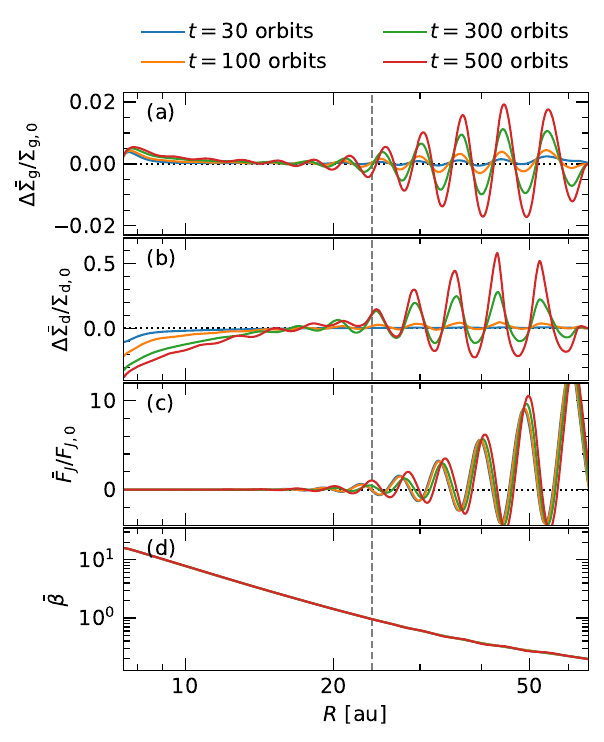}
	\caption{Azimuthally averaged profiles of the perturbed gas and dust surface densities (panels \emph{a} and \emph{b}), the angular momentum flux $F_J$ (panel \emph{c}), and the cooling timescale $\beta$ (panel \emph{d}) as a function of radius and for several snapshots. The AMF is approximately constant over time, leading to deeper gaps and sharper features as the disk evolves. The disk is roughly separated into a featureless inner region and a ringed outer region at $R\approx 24$\,au, which corresponds to $\beta\approx\beta_\text{crit}$ (see Eq.~\eqref{eq:bcrit}). The AMF is arbitrarily normalized to $F_{J,0}=10^{-10}\,\Msun\,\text{au}^2/\text{yr}^2$.}
	\label{fig:fiducial-1D}
\end{figure}
where all quantities refer to the gas and bars denote azimuthal averaging. We then plot azimuthally averaged profiles of the perturbed gas and dust surface densities as well as the AMF $F_J$ as a function of time in Fig.~\ref{fig:fiducial-1D}. This figure confirms that an AMF is indeed present in the disk, with the troughs in the gas surface density corresponding to peaks in the AMF. Dust features then form at the radial locations where the radial pressure gradient is minimized, slightly offset from the peaks in gas surface density due to the radial temperature gradient.

\begin{figure}
	\includegraphics[width=\columnwidth]{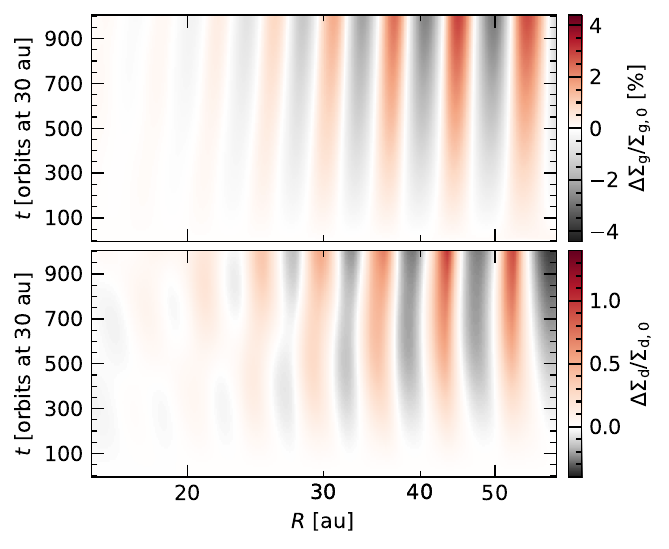}
	\caption{Time evolution of the azimuthally averaged perturbed gas and dust surface densities. Gaps deepen and dust accumulates in rings roughly linearly over time.}
	\label{fig:fiducial-time}
\end{figure}
The fact that $F_J$ remains roughly constant over time also suggests that the gap depth will approximately increase linearly with time, resulting in sharper features as the disk evolves. We verify this statement by plotting the azimuthally averaged perturbed gas and dust densities as a function of time in Fig.~\ref{fig:fiducial-time}, where we show that the undulations in both quantities increase in amplitude over time.

\begin{figure}
	\includegraphics[width=\columnwidth]{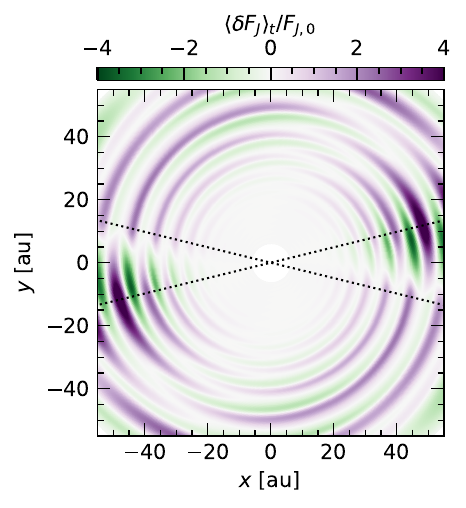}
	\caption{Heatmap of the quantity $\delta F_J$ averaged over 200 snapshots between $t\in[100,300]\,P_0$. A high AMF is present at the exits of the shadows, indicating the deposition of angular momentum into the disk and leading to the sustained formation of gaps. Similar to Fig.~\ref{fig:fiducial-1D},  $F_{J,0}=10^{-10}\,\Msun\,\text{au}^2/\text{yr}^2$.}
	\label{fig:fiducial-amf}
\end{figure}
We also plot a heatmap of the twodimensional quantity $\delta F_J$ averaged over 200 snapshots between $t\in[100,300]\,P_0$ in Fig.~\ref{fig:fiducial-amf}. We find that the AMF is strongest at the exit of either shadow, which also implies that gaps should be slightly deeper at those locations compared to the illuminated regions a quarter of an orbit later. This exact behavior can be seen in the top panels of Fig.~\ref{fig:fiducial-spiral-ring}, with perturbations being strongest at the exits of the shadows and weakest after an angular distance of $\pi/2$.

\subsection{A criterion for the formation of rings}

The dependence of the azimuthal temperature perturbation on the cooling timescale (see Fig.~\ref{fig:fiducial-T}) might hint at the reason why we do not observe rings or spirals in the inner disk. In particular, smaller azimuthal temperature perturbations would result in weaker spiral waves and therefore a smaller AMF, in turn leading to less pronounced (if at all present) rings. By combining the contributions to cooling due to thermal emission from the disk surfaces and in-plane radiative diffusion, we define the cooling timescale $\beta$ following \citet{ziampras-etal-2024b} as
\begin{equation}
    \label{eq:beta}
    \beta = \frac{1}{f+1}\frac{e}{|\Qcool|}\OmegaK,\qquad f=16\pi \frac{\taueff}{\tauR}\frac{\tauP^2}{6\tauP^2+\pi}.
\end{equation}
We then plot the azimuthally averaged cooling timescale $\bar{\beta}$ at $t=0$ in the bottom panel of Fig.~\ref{fig:fiducial-1D}, showing that it decreases with radius as expected.

To crudely estimate the conditions under which temperature perturbations due to the shadows are too small to efficiently generate spiral waves, we require that the cooling timescale $\beta$ becomes longer than the time it takes to cross the shadowed regions \citep[see also,][]{casassus-etal-2019}. Assuming Keplerian rotation, and since we model two shadows, this critical timescale is then
\begin{equation}
	\label{eq:bcrit}
	\beta_\text{crit} \sim 2 \frac{\Delta t_\text{shadow}}\OmegaK \approx 2\frac{2 w_\text{sh} \OmegaK}{\OmegaK} = 4 w_\text{sh},
\end{equation}
or $\beta_\text{crit}\approx 0.96$ for our models, which translates to $R\approx24$\,au. We mark this radial location with a vertical dashed line in Fig.~\ref{fig:fiducial-1D}, showing that it roughly divides the disk into a feature-poor inner region and a ringed outer region, consistent with our expectations.

We note that this is only a rough estimate in the interest of providing a simple explanation for the lack of features in the inner disk. In principle, a weak AMF will exist even for small temperature perturbations due to a shadow, and rings would still form after an arbitrarily long time, which might of course be longer than the lifetime of the disk.

\subsection{Spacing between rings}
\label{sub:spacing}

\begin{figure}
	\includegraphics[width=\columnwidth]{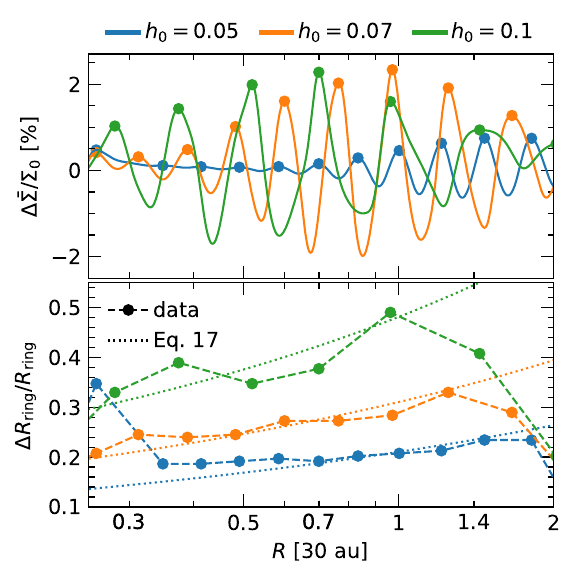}
	\caption{Top: radial profile of the perturbed gas surface density at $t=\{100,30,15\}\,P_0$ for $h=\{0.05,0.07,0.1\}$, respectively. Dots mark the location of peaks. Bottom: radial spacing between rings in the gas surface density as a function of radius. The dotted line is based on Eq.~\eqref{eq:ring-spacing} for $f=2/7$.}
	\label{fig:spacing}
\end{figure}

Figures~\ref{fig:fiducial}~\&~\ref{fig:fiducial-1D} hint at a regular spacing between rings in the gas surface density. We compute this spacing as the radial distance between two adjacent rings normalized to the radius of the interior ring in Fig.~\ref{fig:spacing}, confirming that the rings are fairly regularly spaced in radius.

Given that the disk exhibits two large-scale, continuous spirals (see bottom panels of Fig.~\ref{fig:fiducial-spiral-ring}) due to the two shadows, and considering that the AMF peaks at the shadow exits, we can make a rough estimate of the radial spacing between two rings. To do this we define the pitch angle $\phi_\text{p}$ of the spiral waves as
\begin{equation}
	\label{eq:pitch-angle}
	\tan\phi_\text{p} = \frac{1}{R}\DP{R}{\phi} = \frac{1}{R}\DP{R}{t}\DP{t}{\phi} \approx \frac{\sqrt{\gamma}\cs}{R\OmegaK} = \sqrt{\gamma}h,
\end{equation}
assuming that the waves travel at the adiabatic sound speed $\sqrt{\gamma}\cs$. With $h=h_0\,(R/R_0)^f$, we then integrate Eq.~\eqref{eq:pitch-angle} from one shadow exit to the next to obtain:
\begin{equation}
	\label{eq:ring-spacing}
	\int\limits_{R_0}^{R_0+\Delta R} \frac{1}{hR}\,\mathrm{d}R = \sqrt{\gamma}\int\limits_{\phi_0}^{\phi_0+\pi}\mathrm{d}\phi \Rightarrow \frac{\Delta R}{R_0} = \left(1-\sqrt{\gamma}\pi fh_0\right)^{-1/f} - 1.
\end{equation}
We then plot this result in Fig.~\ref{fig:spacing} with a dotted line for $f=2/7$ and all values of $h$ (discussed further in Sect.~\ref{sub:aspect-ratio}), showing that this simple model captures the spacing between rings fairly well, aside from the left and right edges of the disk due to radial boundary effects.

\section{Parameter study}
\label{sec:parameter-study}

Having analyzed the fiducial model, we explore here the effects of varying the coagulation fraction $X$, the viscosity $\alpha$, and the aspect ratio $h$.

\subsection{Coagulation fraction $X$}
\label{sub:coagulation-fraction}

\begin{figure}
	\includegraphics[width=\columnwidth]{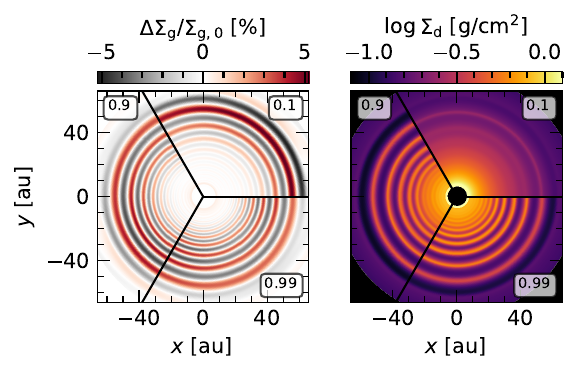}
	\caption{Heatmaps similar to Fig.~\ref{fig:fiducial} for coagulation fractions $X=\{0.1,0.9,0.99\}$ at $t=1000\,P_0$. The $X=0.1$ model is dominated by small grains and is therefore optically thicker, resulting in rings forming slightly further out due to the cooling criterion in Fig.~\eqref{eq:bcrit}. The rings are also much fainter in this model due to the smaller fraction of big grains.}
	\label{fig:vary-X}
\end{figure}

Figure~\ref{fig:vary-X} shows heatmaps of the perturbed gas density and the dust density similar to Fig.~\ref{fig:fiducial} for the three different values of the coagulation fraction $X\in\{0.1, 0.9, 0.99\}$ used in our study. Here, we find that the effect of having a significant fraction of the dust mass as small grains ($X=0.1$) is twofold. For one, it results in rather faint and less sharp features in the dust distribution, as the majority of the dust is well-coupled to the gas and does not accumulate on the forming pressure bumps.

\begin{figure}
	\includegraphics[width=\columnwidth]{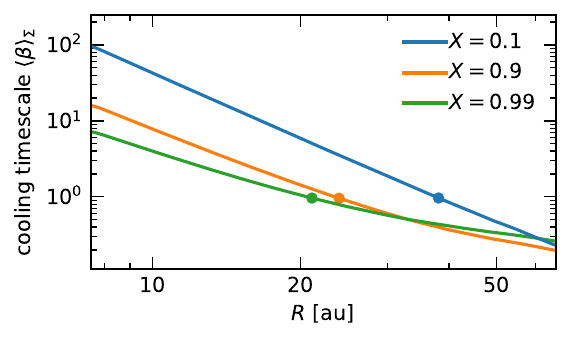}
	\caption{The azimuthally averaged, density-weighted cooling timescale following Eq.~\eqref{eq:beta} at $t=0$ for the three models with different $X$. A lower $X$ results in more small grains and an optically thicker, slowly-cooling disk. A dot marks the radial location where $\beta\approx\beta_\text{crit}$ according to Eq.~\eqref{eq:bcrit}.}
	\label{fig:beta-X}
\end{figure}

At the same time, the higher opacity of the submicron-sized grains renders a larger portion of the inner disk optically thick, resulting in slower cooling and ultimately limiting the formation of rings in the gas (and therefore the dust) structure. To verify this statement, we show a radial profile of the cooling timescale at $t=0$ for all three models following Eq.~\eqref{eq:beta} in Fig.~\ref{fig:beta-X}. Through this figure we can also explain why the models with $X\geq0.9$ look quite similar: the cooling timescale is quite similar between the two for $R\gtrsim 20$\,au.

Nevertheless, for realistic values of $X\gtrsim 0.9$ that account for the dust growth happening in a typical protoplanetary disk by its T-Tauri stage \citep[e.g.,][]{birnstiel-2023}, we find that cooling is sufficiently fast to facilitate the formation of rings in the bulk of the disk.

\subsection{Turbulent $\alpha$}
\label{sub:turbulence}

\begin{figure}
	\includegraphics[width=\columnwidth]{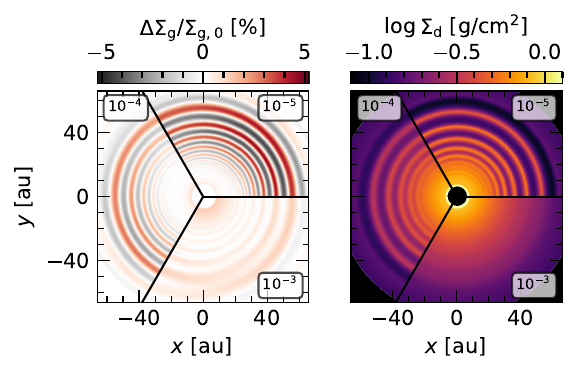}
	\caption{Similar to Fig.~\ref{fig:vary-X}, for different viscosity parameters $\alpha=\{10^{-5},10^{-4},10^{-3}\}$. Rings are slightly more diffuse for $\alpha=10^{-4}$ and hardly visible for $\alpha=10^{-3}$ when compared to the fiducial model with $\alpha=10^{-5}$.}
	\label{fig:vary-alpha}
\end{figure}

The effect of modeling turbulence via a viscous $\alpha$ stress is rather straightforward, acting against the formation of radial structure by diffusing the ring-like gas features and related pressure bumps. It is nevertheless useful to check how strong turbulence should be to eliminate (or at least strongly damp) the substructures we find in our fiducial model.

In Fig.~\ref{fig:vary-alpha} we show our results for models with $\alpha=10^{-4}$ and $10^{-3}$, complementing our fiducial value of $10^{-5}$. While the model with $\alpha=10^{-4}$ shows slightly more diffuse rings with a weaker contrast compared to our fiducial case, the radial structure is still clearly visible. This is not the case for $\alpha=10^{-3}$, however, where the viscous torques manage to wash away any developing radial features.

Overall, for typical values of $\alpha\lesssim10^{-4}$ motivated by hydrodynamical mechanisms such as the vertical shear instability \citep[e.g.,][]{nelson-etal-2013,flock-etal-2020b} and the lack of efficient magnetohydrodynamical turbulence in the ``magnetically dead'' zone of the disk \citep{bai-stone-2013}, we expect that turbulence should not inhibit significantly the formation of rings through the mechanism studied in this paper.

\subsection{Aspect ratio $h$}
\label{sub:aspect-ratio}

\begin{figure*}
	\includegraphics[width=\textwidth]{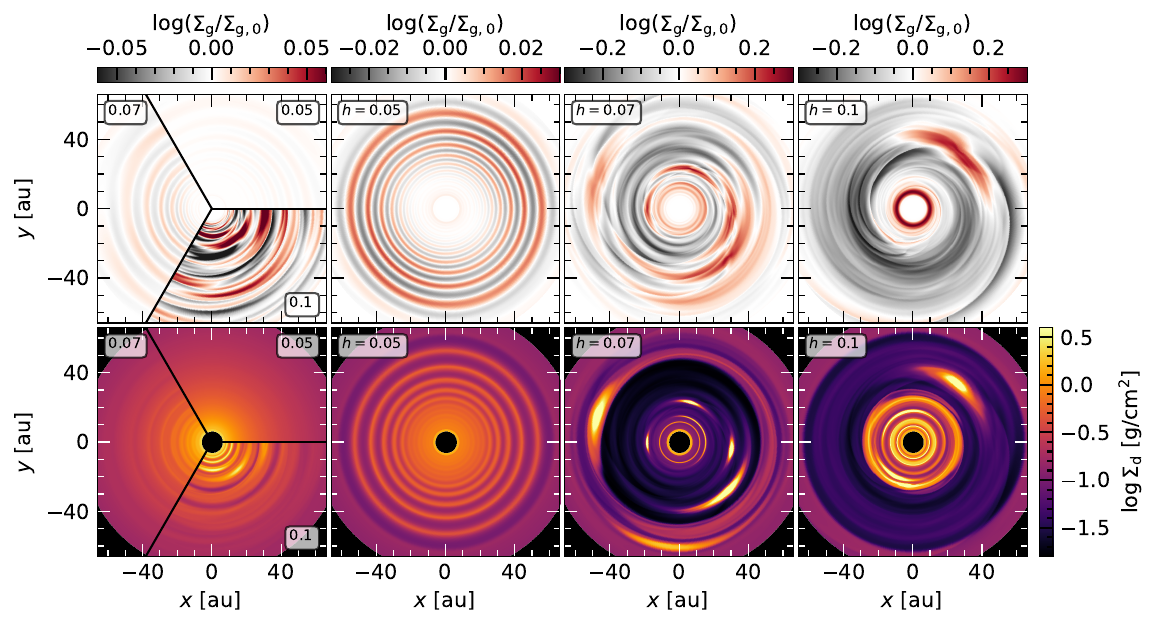}
	\caption{Top: the perturbed gas surface density for different aspect ratios, marked in the corner of each panel. The leftmost panel corresponds to $t=50\,P_0$, while the others refer to $t=1000\,P_0$. Bottom: the dust surface density for the same models. The significantly shorter cooling timescale for larger $h$ drives a rapid formation of deep gaps, which often result in the nearby rings collapsing into vortices due to the RWI.}
	\label{fig:panels-h}
\end{figure*}

Given that the aspect ratio $h$ largely determines the width of the shadow through the scale height of the misaligned inner disk (see Fig.~\ref{fig:setup}), the cooling timescale as $\beta\propto h^{-6}$ (see Eq.~\eqref{eq:beta}), and the radial spacing between rings (Eq.~\eqref{eq:ring-spacing}), it is very interesting to investigate how the disk structure changes with $h$ even though we hold the shadow width $w_\mathrm{sh}$ fixed. We show a comparison among models with $h(R_0)=\{0.05,0.07,0.1\}$ in Fig.~\ref{fig:panels-h}.

Regarding the formation of rings, a higher $h$ results in a significantly shorter cooling timescale, which in turn leads to a higher temperature contrast between the shadowed and illuminated regions (see Fig.~\ref{fig:fiducial-T}), and therefore more vigorous spiral waves, a higher angular momentum flux into the disk, and ultimately more efficient gap opening. This is clearly visible on the left panels of Fig.~\ref{fig:panels-h}, which show snapshots at $t=50\,P_0$ and illustrate the rapid formation of structure in both gas and dust for higher $h$. In particular, gap opening is so efficient for $h=0.1$ that the gap edges are already unstable to the Rossby wave instability \citep[RWI,][]{lovelace-1999} after only $50\,P_0$, resulting in the formation of vortices in the gas surface density and bright clumps in the dust.

Within $1000\,P_0$, the disks with $h(R_0)\geq0.07$ destabilize to the RWI throughout, developing a series of vortices that span the full radial extent of the disk. These vortices decay as they orbit, in part due to viscous diffusion, cooling \citep{rometsch-etal-2021,fung-ono-2021} and dust--gas coupling \citep{raettig-etal-2015, lovascio-etal-2022, ziampras-etal-2024b}. For $h=0.1$, several vortices in the outer disk ($R\gtrsim20$\,au) merge into a single large vortex, while most of the vortices that formed in the inner disk eventually dissipate along the azimuthal direction, resulting in the dust structure reverting to a set of rings. 

The vortex-rich structure seen in models with $h(R_0)\geq0.07$ is in stark contrast to the fiducial model with $h=0.05$, which shows a purely radial structure. We highlight, however, that the gap opening process is still ongoing and will inevitably lead to gap edges steep enough to be unstable to the RWI even for this model. Whether this will happen within the lifetime of the disk is beyond the scope of this study.

% \begin{figure}
% 	\includegraphics[width=\columnwidth]{deltaT-all.pdf}
% 	\caption{a}
% 	\label{fig:deltaT-all}
% \end{figure}

\section{Synthetic observations}
\label{sec:observations}

In this section we test whether the features found in our numerical models could be observable with current observational facilities such as the Atacama Large Millimeter Array (ALMA) and the Very Large Telescope (VLT). To that end, we show synthetic observations of the disk models presented in the previous section using the radiative transfer code \radmc{} \citep{dullemond-etal-2012}. We focus on the dust continuum emission at 1.3\,mm (231 GHz, corresponding to ALMA Band~6) and the scattered-light emission at 1.65\,$\mu$m (corresponding to VLT-SPHERE in H~band) to examine the disk structure of the midplane and surface layers of the disk, respectively.

\subsection{\radmc{} setup}
\label{sub:radmc-setup}

Since we operate in a vertically integrated framework for our suite of hydrodynamical simulations, we construct a 3D model of the disk assuming a vertically isothermal disk in hydrostatic equilibrium. We follow a similar procedure as in Appendix~\ref{apdx:two-disks}, this time replacing the surface density profile of the outer disk with that from our hydrodynamical models within the radial range $R\in[0.2,2.5]\,R_0 = [6,75]$\,au. The vertical dust profile for both big and small grains is computed assuming a vertically constant turbulent mixing parameter $\alpha_z$ following \citet{fromang-nelson-2009}
\begin{equation}
	\label{eq:dust-vertical}
	\rho_\text{d}^i = \frac{\Sigma_\text{d}^i}{\sqrt{2\pi}H} \sqrt{\frac{\St_\text{mid}^i}{\alpha_z}+1}\,\exp\left\{-\frac{\St_\text{mid}^i}{\alpha_z} \left[\exp\left(\frac{z^2}{2H^2}\right)-1\right] - \frac{z^2}{2H^2}\right\},
\end{equation}
where $\St_\text{mid}^i$ is the Stokes number of species with index $i$ evaluated at the midplane through Eq.~\eqref{eq:stokes-number} and $H$ is the pressure scale height of the gas. The wavelength-dependent opacity table for each grain species is computed using \optool{} using the composition in Sect.~\ref{sub:parameters}.

Even though the hydrodynamical simulations processed with \radmc{} correspond to runs with $\alpha=10^{-5}$, we choose $\alpha_z=10^{-4}$ for our \radmc{} models. This choice is motivated by the fact that turbulent diffusion does not affect the ring formation process significantly at $\alpha=10^{-4}$ (see Fig.~\ref{fig:vary-alpha}), while a higher $\alpha_z$ helps recover the temperature structure of the disk without the need for an extremely large number of photons.

\begin{figure}
	\includegraphics[width=\columnwidth]{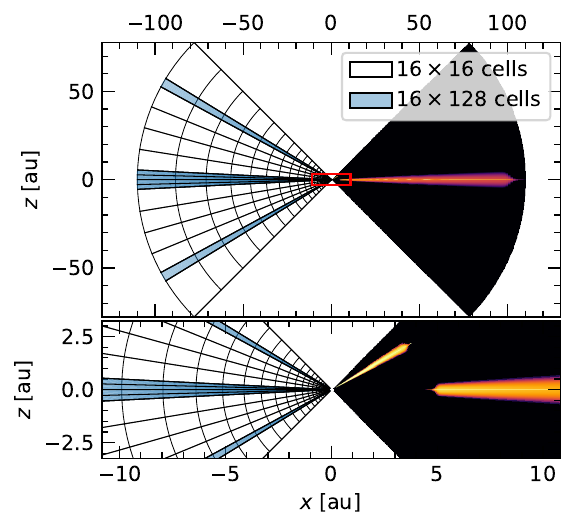}
	\caption{Left: vertical cross-section of the grid used in \radmc{} models. Each cell in the figure corresponds to a block of cells in the grid used, with the color indicating the number of cells in the block. Right: dust density structure for $h_0=0.05$. The bottom row zooms into the red box in the top row.}
	\label{fig:radmc-grid}
\end{figure}

To resolve the highly settled layer of mm grains in the polar direction for both the inner misaligned disk and the outer disk, we use a grid with
\begin{itemize}
	\item 256 cells between $\theta\in\left[\frac{\pi}{2}, \frac{\pi}{2}+h_0\right]$,
	\item 64 cells between $\theta\in\left[\frac{\pi}{2}+h_0, \theta_\text{tilt}
	-\frac{h_0}{2}\right]$, $\theta_\text{tilt} = \frac{\pi}{2}+30^\circ$,
	\item 128 cells between $\theta\in\left[\theta_\text{tilt}-\frac{h_0}{2}, \theta_\text{tilt}+\frac{h_0}{2}\right]$, and
	\item 16 cells between $\theta\in\left[\theta_\text{tilt}+\frac{h_0}{2}, \pi\right]$,
\end{itemize}
with a mirror symmetry about the midplane $(\theta=\pi/2)$ for a total of $N_\theta=920$ cells. The radial extent covers $r\in[0.1,110]$\,au with $N_r=512$ logarithmically spaced cells, and the azimuthal domain spans $\phi\in[0,2\pi]$ with $N_\phi=514$ cells. A cross-section of the grid used in \radmc{} is shown in Fig.~\ref{fig:radmc-grid}. The dust surface density from the hydrodynamical models is first interpolated onto the midplane of the 3D grid $\{r,\theta=\pi/2,\phi\}$, and the 3D density structure for each species is then computed using Eq.~\eqref{eq:dust-vertical}.

To calculate the dust temperature with \radmc{} we use the \texttt{mctherm} command with $10^{10}$ photons and a solar-type star ($\Mstar=1\,\Msun$, $T_\star=5772$\,K). To make sure that the frequency range for stellar emission is the same for all values of $h_0$, we change the stellar radius to $R_\star=[1, 3.25, 11.34]\,R_\odot$ such that $\Lstar=[1, 10.6, 128.7]\,\Lsun$ for $h_0=[0.05, 0.07, 0.1]$, respectively, and treat the star as a point source to avoid unexpected geometrical effects. This is a good approximation for our setup, as $R \gg R_\odot$ even for the inner disk. The vertical structure of the misaligned inner disk is also now computed with the appropriate $h_0$ such that its shadow can be captured in the dust temperature calculation. Finally, we include the effects of anisotropic scattering using the approximation of \citet{henyey-greenstein-1941} with the scattering asymmetry parameter $g$ computed using \optool{}.

It is important to note here that the temperatures computed by \radmc{} are equilibrium temperatures, in contrast to those used in the dynamic calculations. As a result, \radmc{} cannot capture the asymmetric azimuthal variation of temperature found in the hydrodynamical models around the shadow. We highlight this difference in Appendix~\ref{apdx:azimuthal-temperature}. Given that the azimuthally averaged midplane temperature profiles obtained from \radmc{} are remarkably similar to those from the hydrodynamical models, we replace the \radmc{} output of the temperature structure within $\pm H/2$ about the midplane with the hydrodynamical temperature structure. In doing so, our models can also capture the azimuthal variations in temperature (see e.g., Fig.~\ref{fig:fiducial-T}) when computing synthetic images in mm emission. We believe that applying this correction to the disk surface would not necessarily produce more accurate results, as our models are 2D, but instead refer the reader to \citet{zhang-zhu-2024}, who computed scattered light images for their (3D) models.

We then compute images at 1.3\,mm and 1.65\,$\mu$m with the \texttt{image} command for various orientations using $10^8$ photons and 768 pixels on both $x$ and $y$ directions, assuming a distance of 100\,pc to the source. For images at 1.65\,$\mu$m, the flux within a circle of radius 100 milliarcseconds (mas) from the star is blocked to emulate the use of a coronagraph. The images are then convolved with a Gaussian beam of 40\,mas for both wavelengths to mimic typical observations with ALMA Band 6 and SPHERE, respectively.

\subsection{Synthetic images}
\label{sub:synthetic-images}

\begin{figure*}
	\includegraphics[width=\textwidth]{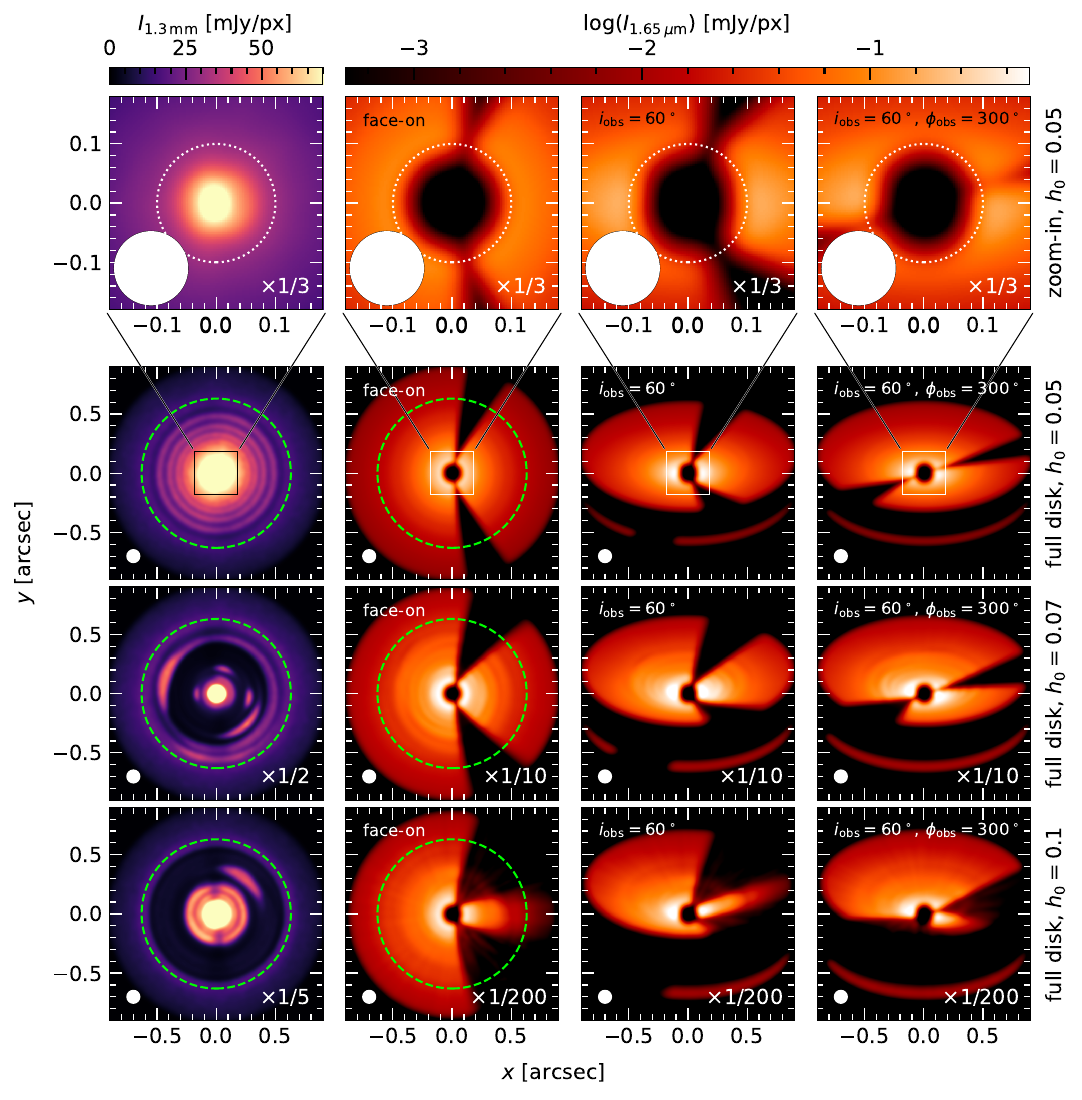}
	\caption{Synthetic images for our fiducial model with $h_0=0.05$ (top two rows) as well as $h_0=0.07$ and $h_0=0.1$ (third and fourth row) at 1.3\,mm (leftmost column) and 1.65\,$\mu$m (remaining panels) for different disk inclinations and azimuthal rotations. The top row zooms into the (unresolved) inner region at $R\lesssim18$\,au, marked by a box in each panel of the second row. Rings and shadows are visible throughout the entire disk for all values of $h_0$. For better visibility on the same scale, the fluxes are rescaled by a factor shown on the bottom right corner of each panel. A solid white circle marks the beam size of 40~mas, and a dashed white circle the coronagraph of radius 100\,mas. A dashed green circle marks the outer edge of our hydrodynamical domain.}
	\label{fig:radmc-gallery}
\end{figure*}

In the top two panels of Fig.~\ref{fig:radmc-gallery} we present the synthetic images for our fiducial model ($h_0=0.05$, $X=0.9$, $\alpha=10^{-5}$, $\alpha_z=10^{-4}$) after $1000\,P_0$. The left panels show the dust continuum emission at 1.3\,mm assuming a face-on disk, while the remaining panels show the scattered-light emission at 1.65\,$\mu$m for different inclinations and azimuthal rotations of the disk with respect to the observer.

In the second row of Fig.~\ref{fig:radmc-gallery}, which shows the full disk structure, the dust continuum emission features both a series of bright rings as well as a faint azimuthal variation in brightness due to the shadow cast by the misaligned inner disk. In the remaining panels of the same row, corresponding to the scattered-light images, the shadowed wedge is clearly visible but the rings are not. The top row of the same figure shows zoomed-in views of the inner disk regions, highlighting the directly illuminated inner rim of the outer disk ($R\lesssim18$\,au). However, the structure of this inner region is blurred due to its angular size being comparable to the beam size. 

In the third and fourth row of Fig.~\ref{fig:radmc-gallery} we show the synthetic images for models with $h_0=0.07$ and $h_0=0.1$, respectively, similar to the second row of the same figure. As expected, the vortices that form in both models are clearly visible as bright arcs in mm emission, and can still be faintly observed in scattered light, but the rings are too faint to be observable. This is a result of the very weak contrast in $\mu$m-sized grains, which largely trace the gas and therefore show $\Delta\Sigma/\Sigma\lesssim 5\%$. The shadows grow progressively wider with increasing $h_0$ due to the larger scale height of the inner disk, and a dark lane can even be observed in mm emission due to the stark temperature contrast between the shadowed and illuminated regions.

Overall, we find that the features in our hydrodynamical models should be observable with current facilities and especially so for the models with larger aspect ratios, as they both exhibit more pronounced substructure and represent hotter, brighter disks. For the case of our fiducial model for $h=0.05$, however, the features are fainter and might be harder to detect. For this reason, we further process the (unconvolved) images at 1.3\,mm using the \simio{} tool \citep{kurtovic-2024}\footnote{\simio{} utilizes the \casa{} package \citep{casa-team-2022}.} to simulate an ALMA observation that more accurately represents the noise induced by the synthesized ALMA beam. We use a configuration that corresponds to the observation of a source analogous to the system HD~163296 \citep[$d=101$\,pc, $i=47^\circ$, $\text{P.A.}=133^\circ$, see e.g.,][]{huang-etal-2018}.

\begin{figure}
	\includegraphics[width=\columnwidth]{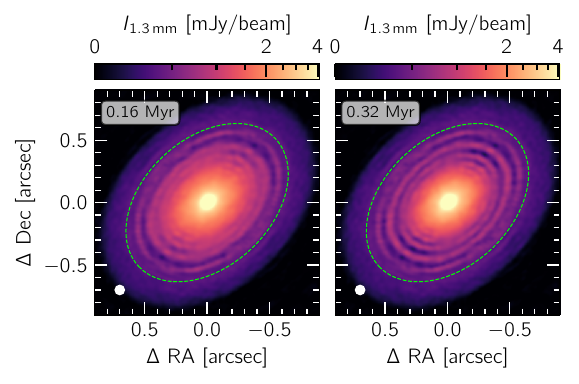}
	\caption{Synthetic ALMA image at 1.3\,mm for our fiducial model (see second row of Fig.~\ref{fig:radmc-gallery}), configured to mimic an observation of the system HD~163296 using \simio{}. Left: $t=1000\,P_0\approx0.16$\,Myr. While faint, rings and gaps are visible by eye. Right: $t=2000\,P_0\approx0.32$\,~Myr. The rings are now clearly visible as gap opening progresses. Colors are rescaled with an arcsinh stretch.}
	\label{fig:simio-h05}
\end{figure}

Figure~\ref{fig:simio-h05} shows the synthetic ALMA image for our fiducial model with $h_0=0.05$, with the left panel corresponding to $t=1000\,P_0$ as in all panels shown so far in this section. While the contrast is rather weak, rings and gaps are visible to the naked eye without requiring any post-processing. The shadow cast by the misaligned inner disk is, however, not discernible by eye. To highlight the time-sensitive nature of the gap opening process, we also show the synthetic ALMA image at $t=2000\,P_0$ in the right panel of the same figure, where the rings are now clearly visible as gap opening progresses and more dust accumulates in the resulting pressure bumps.

\begin{figure}
	\includegraphics[width=\columnwidth]{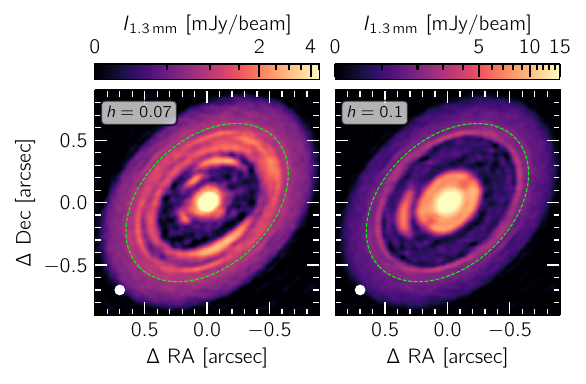}
	\caption{Same as Fig.~\ref{fig:simio-h05} for models with $h_0=0.07$ (left) and $h_0=0.1$ (right) at $t=1000\,P_0$. Vortices are clearly visible, but fainter features such as rings and arcs corotating with the vortices are less discernible by eye.}
	\label{fig:simio-h07-10}
\end{figure}

For completion, we also show the synthetic ALMA images for the models with $h_0=0.07$ and $h_0=0.1$ in Fig.~\ref{fig:simio-h07-10}. In these cases, vortices are clearly visible for both models, but the arcs of dust corotating with the vortices for $h_0=0.07$ as well as the rings interior to the vortex for $h_0=0.1$ are no longer visible by eye. For $h_0=0.1$, a dark lane due to the shadow cast by the misaligned inner disk is clearly visible in the synthetic ALMA image, as the shadow is wider and more pronounced for larger aspect ratios.

\section{Discussion}
\label{sec:discussion}

In this section we discuss the implications of our results in the context of planet formation and observations of protoplanetary disks. We also address additional effects that we have not considered in our study, but can complicate the picture.

\subsection{Implications for planet formation}
\label{sub:planet-formation}

Recent years have shown that substructures in protoplanetary disks appear to be ubiquitous \citep{Andrews2018,Long2018}. They are of crucial importance for planet formation models \citep{Guilera2020,Morbidelli2020}, as they are trapping pebbles and thereby facilitating the growth of planetesimals towards planets \citep{Lau2022,Jiang2023}. Planets, however, are one of the most popular mechanisms to explain the observed substructures as they are indicated by kinematics \citep[e.g.][]{Teague2018,Pinte2018,Teague2019} or in exceptional cases even directly detected \citep{Keppler2018}. Planets and substructures may in fact trigger each other, as demonstrated in \citet{Lau2024}. It is therefore crucial to understand which effects can lead to the formation and evolution of substructures in disks and how to distinguish between the various mechanisms.

In this work, the shadows cast by an inclined inner disk lead to the formation of multiple rings. This may play a significant role in this process as it can seed an outer disk with structures that kick-start planet formation through dust accumulation. However, with numerous rings, each may only trap a limited amount of dust, potentially hindering efficient planet formation if the material is too spread out. The impact of these shadow-induced rings likely evolves over time, as the disk cools and the dust population changes. The demonstrated formation of vortices as final stage of the ring evolution can further accumulate dust also along the azimuthal direction. This time evolution could help explain why we observe relatively few substructures in younger Class 0/I disks, but far more ringed structures in later stages of disk evolution.

\subsection{Effect of a moving shadow}
\label{sub:moving-shadow}

In our models we assume that the shadow is static in time, or equivalently that the misaligned inner disk is not precessing. This choice was made for the sake of simplicity, although misaligned inner disks are expected to precess \citep[e.g.,][]{facchini-etal-2018,kuffmeier-etal-2021}, with a common explanation being an inclined planet \citep[e.g.,][]{nealon-etal-2018} or binary companion \citep[e.g.,][]{facchini-etal-2018}. A shadow due to a warp in the inner region of the disk \citep[e.g.,][]{papaloizou-pringle-1983} would also precess \citep{kimmig-dullemond-2024}. If the warp propagates too fast to be eliminated by viscous damping, it could result in a broken inner disk \citep{dougan-etal-2015}, which would then precess as well.

In the case of a precessing shadow, \citet{montesinos-cuello-2018} showed that the emerging spiral structure resembles planet-like signatures, with the radial location of the spiral-launching point being a function of the precession direction and frequency of the shadow. A radial profile of the cooling timescale complicates the picture, as forming radial substructure would require an appropriate balance between the cooling, shadow precession, and orbital/shadow crossing timescales.

\subsection{Comparison to related studies}
\label{sub:su-bai-2024}

\citet{cuello-etal-2019} were first to investigate the trapping of dust particles in spirals driven by shadows in a framework similar to ours, but their models did not feature rings or gaps. Instead, the spiral structures presented in their work are much brighter in 1.3\,mm emission, contrary to our models where the spirals are quickly overshadowed by the forming rings or vortices. All of the above can be attributed to several different reasons.

For one, the omission of in-plane radiative diffusion in \citet{cuello-etal-2019} renders the disk less radiatively efficient, as the in-plane cooling channel (which can be much more efficient in the presence of temperature gradients) is unavailable \citep{miranda-rafikov-2020b,ziampras-etal-2023a}. This results in a longer cooling timescale and therefore slower development of the gap opening process. In combination with their significantly shorter integration timescale of $10^4$\,yr---10~orbits at 100\,au, their reference radius---it is not surprising that the only prominent feature in their models is the spiral structure, which develops very early on.

Furthermore, the brightness of the spirals in mm emission, which primarily depends on the local dust-to-gas ratio, heavily depends on the background dust model. \citet{cuello-etal-2019} used a dust size distribution with $a_\text{d}^\text{max}=10$\,cm, or $\St^\text{max}\sim2$ at 100\,au, such that a substantial fraction of dust grains lies in the regime $\St\approx0.1$--1 and can very efficiently accumulate around the spiral arms. The fact that the dust particles were treated using post-processing can further exaggerate this effect, as dust--gas dynamics are missed. In contrast, our models feature mm grains with $a_\text{d}^\text{big}=1$\,mm or $\St\approx0.01$ at $R_0=30$\,au, which do not experience such efficient accumulation around spirals but will nevertheless be trapped in the resulting pressure bumps.

Nevertheless, the studies by both \citet{montesinos-etal-2016} and \citet{cuello-etal-2019} laid the groundwork in probing the observational signatures of spirals induced in shadowed disks, and in particular our results relate to \citet{cuello-etal-2019}, where the central focus was the dynamics of the dust component. Our results, while seemingly qualitatively different, are in fact fully consistent with theirs once the differences in parameter space and model complexity are taken into consideration.

During the preparation of this manuscript, we became aware of the work by \citet{su-bai-2024} (hereafter SB24), who also studied the dynamical evolution and formation of substructure in shadowed protoplanetary disks. SB24 opted for a more controlled, statistical approach by using a simplified, constant cooling timescale $\beta$ per model and carrying out a large set of simulations for different $\alpha$, $w_\text{sh}$, $h$, and $\beta$. They then commented on the existence of different ``branches'' in the parameter space, with the formation of rings being more likely for disks with $\beta\sim1$ and a weak temperature contrast in the shadow $|\Delta T|/T\lesssim0.15$, and vortices forming for $\beta=10^{-3}$.

Our results are entirely consistent with their findings, and provide further insight into the physical mechanisms behind the formation of rings and vortices. For a realistic temperature contrast $|\Delta T|/T\sim0.15$ for $h_0=0.05$, all models are firmly in the nonlinear regime where spirals can deposit angular momentum into the disk. It then becomes a question of whether the angular momentum flux is sufficiently high to open gaps, a process mitigated by a high $\alpha$ or limited by the simulation runtime (or, more realistically, the disk lifetime). For sufficiently low $\alpha$, the gap edges are eventually unstable to the RWI, triggering the formation of vortices. In that regard, the three branches (``spirals'', ``rings'', and ``vortices'') presented in SB24 can be understood as different snapshots of the same process.

We also note that their quoted value of $\beta\sim1$ needed to form rings is entirely consistent with our criterion in Eq.~\eqref{eq:bcrit}, which for their models evaluates to $\beta_\text{crit}\approx 4 \sigma_\phi \approx 1$. Given that our models already show vortices by the end of the simulation runtime for $h_0=0.07$, where $\beta\lesssim 0.1$ at $R\gtrsim R_0$, it makes sense that the disk destabilizes to the RWI nearly immediately for $\beta=10^{-3}$ in their models and further suggests that a cooling timescale of $\beta\lesssim0.1$ should be sufficient to trigger the RWI well within the disk lifetime under the constant thermal forcing of a shadow.

In addition, recently \citet{zhang-zhu-2024} showed with threedimensional radiative hydrodynamics simulations that the spiral-driving mechanism due to a shadow is robust and can excite substantial angular momentum transport as well as bright features in scattered-light observations. While their work focused on the spiral formation process and its observable signatures, it is worth pointing out that their models also featured gaps and vortices (see Fig.~2 therein) due to the same mechanism studied in this work. Combined with the recent linear theory calculations by \citet{zhu-etal-2024}, it is clear that shadow-induced spirals can very efficiently transport angular momentum and drive substructure in protoplanetary disks.

All in all, our findings are in very good agreement with previous and current studies, with a key difference being the inclusion of dust hydrodynamics. In this context, the latter enables more consistent dust--gas interaction both via momentum exchange and the determination of the opacity, as well as synthetic observations of mm emission with ALMA. Further work is needed to understand dust--gas dynamics in a threedimensional model, given that the excitation of spirals can drive substantial vertical motion \citep{zhang-zhu-2024} which can set the dust scale height and therefore overall cooling efficiency.

\subsection{Observational prospects}
\label{sub:observations}

In Sect.~\ref{sec:observations} we showed that the rings formed due to the shadowing by a misaligned inner disk should be visible with modern observational facilities such as ALMA and VLT-SPHERE. At the same time, however, the shadow itself might not be detectable for disks with a larger aspect ratio (i.e., hotter). It is therefore possible that several observed disks may exhibit substructure due to a shadow, but the shadow itself might not be visible. Nevertheless, systems with both a visible shadow and substructure have been observed, making them ideal targets for further study. Such systems include HD~100453 \citep{benisty-etal-2017}, HD~142527 \citep{avenhaus-etal-2017}, and HD~143006 \citep{benisty-etal-2018,andrews-etal-2018}.

In addition, in Fig.~\ref{fig:spacing} we showed that the radial spacing between rings is highly regular and depends solely on the aspect ratio of the disk, a quantity that can be constrained with observations of gas tracers at different altitudes from the disk midplane \citep[e.g.,][]{law-etal-2021}. This regular spacing could be used as a diagnostic tool to infer whether the observed rings are due to a shadow or another mechanism, such as magnetic fields \citep{bethune-etal-2017,riols-etal-2020} or undetected exoplanets \citep[e.g.,][]{zhang-etal-2018}, as both of these mechanisms result in irregularly spaced rings in mm emission. In particular, disks such as around AS~209 \citep{Guzman2018} or SO~1274 \citep{Huang2024} display many narrowly separated rings, which may be linked to the mechanism explored in this work.

Our results also suggest that the formation of substructure is encouraged for short cooling timescales, which is generally true for hotter disks. With that in mind, it is possible that the formation of rings or vortices due to a shadow is more common in disks around Herbig Ae/Be stars. %Interestingly, shadows caused by misaligned inner disks tend to have a higher detection rate around Herbig stars \citep[e.g.,][]{citation-needed} \alexnote{citation needed}, which could be a hint that the formation of substructure due to a shadow is more common in these systems.

Nevertheless, a misalignment of more than $\sim 10^\circ$ is required for the shadow to manifest as a sharp, narrow dark lane rather than an extended shadow. The latter would most likely induce warps at the disk surface rather than strong spirals and rings at the midplane, the dynamical consequences of which might be more difficult to observe.

\subsection{Limitations of this work}
\label{sub:limitations}

In this study we have made several simplifying assumptions. For one, we have worked in a vertically integrated (2D) framework, which neglects both the propagation of waves in the vertical direction and the threedimensional structure of a shadow cast by a misaligned disk. As a result, the dynamical evolution of possible warps \citep[e.g.,][]{kimmig-dullemond-2024} is not captured in our models. Our 2D framework further assumes that the dust is in settling--mixing equilibrium, which might not be a good approximation around the shadowed regions. A fully 3D hydrodynamical model would be required to address these issues, but would also be much more computationally expensive. This will be the focus of followup work.

Furthermore, in our models we assume that the dust and gas grains are perfectly thermally coupled such that cooling is limited by the optical properties of dust grains rather than their collisional coupling to the gas. This assumption can easily break down in the outer disk, and especially so for larger grains \citep[e.g.,][]{dullemond-etal-2022,muley-etal-2023}. In this case, the cooling timescale would be longer than what is shown in Fig.~\ref{fig:beta-X} in the outer disk, possibly inhibiting the formation of rings for $R\gtrsim 70$\,au. We nevertheless carry out a comparison between our radiation hydrodynamical models and a dynamical multi-frequency radiative transfer model in Appendix~\ref{apdx:radiative-transfer}, showing that a large enough temperature contrast between the shadowed and illuminated regions is still present when accounting for the finite thermal coupling timescale between gas and dust species.

Finally, we have neglected magnetic fields, and in particular the role of non-ideal MHD effects, which have been shown to result in the formation of rings \citep{riols-etal-2020}. It is possible that this process could compound with the one studied in this paper, such that the MHD-driven gap opening would accelerate the formation of gaps seeded by the spirals launched at the shadow edges.

\section{Summary}
\label{sec:summary}

In this study we investigated the dynamical evolution of a protoplanetary disk with shadows cast by a misaligned inner disk using high-resolution 2D hydrodynamical simulations. Our models feature a self-consistent treatment of radiation transport including radiative diffusion, and a dust component with dust--gas thermal and dynamical coupling.

We found that the thermal forcing due to the shadowing effect induces a series of spiral waves which deposit angular momentum into the disk at the shadow edges, leading to the formation of concentric gaps that trap mm grains efficiently, in turn forming bright rings in the dust surface density. The gap depth increases linearly with time, as spirals drive a sustained angular momentum flux into the disk, with the process being efficient as long as the local cooling timescale is approximately shorter than the time it takes for gas to cross the shadow. Further analysis showed that the rings are spaced regularly in the radial direction, with their separation depending solely on the disk aspect ratio.

We then carried out a parameter study and found that the formation of rings in the disk is a robust process that is largely insensitive to the coagulation fraction $X$ and the viscous parameter $\alpha$ for reasonable levels of dust growth and turbulence, respectively, as long as the disk is sufficiently optically thin to allow for efficient cooling. The aspect ratio $h$ (as a proxy for the temperature), on the other hand, plays a crucial role in the formation of rings as the cooling timescale depends sensitively on $h$. Models with a large $h$ (i.e., hotter disks) exhibit vigorous spirals and efficient gap opening, leading to Rossby-wave unstable gap edges and the formation of numerous vortices. The latter can appear as arcs, merge into clumps, or diffuse into rings depending on the local thermodynamics and dust--gas interaction.

Following up on our hydrodynamical models, we computed a suite of images of synthetic observations using the radiative transfer code \radmc{} to check whether the features found in our models could be observable with current facilities such as ALMA and VLT-SPHERE. We found that both the rings and the shadow are visible in dust continuum emission at 1.3\,mm for our fiducial model, and a dark lane can be seen in scattered-light images at 1.65\,$\mu$m. The vortices found in models with higher aspect ratios are clearly visible as well, albeit with less pronounced shadows from the misaligned inner disk. Simulated ALMA images show that both the rings and vortices should be detectable with current facilities if present, with the characteristic regular spacing between rings being a diagnostic tool to infer the presence of a shadow.

Our results suggest that shadows can drive substructure in protoplanetary disks in the form of rings, arcs, and vortices, all of which should be observable with current facilities. Dust trapping at the resulting pressure bumps can facilitate the formation of planetesimals via the streaming instability, providing a possible solution to the long-standing chicken-and-egg problem of planet formation.

% \newpage
\section*{Acknowledgments}
The authors thank the anonymous referee for a careful and constructive report that helped identify errors related to the \radmc{} models and greatly improved the quality of the manuscript. AZ would like to thank Xuening Bai, Ondřej Chrenko, Dhruv Muley, Giovanni Rosotti, and Prakruti Sudarshan for helpful suggestions and fruitful conversations. This research utilized Queen Mary's Apocrita HPC facility, supported by QMUL Research-IT (http://doi.org/10.5281/zenodo.438045). This work was performed using the DiRAC Data Intensive service at Leicester, operated by the University of Leicester IT Services, which forms part of the STFC DiRAC HPC Facility (www.dirac.ac.uk). The equipment was funded by BEIS capital funding via STFC capital grants ST/K000373/1 and ST/R002363/1 and STFC DiRAC Operations grant ST/R001014/1. DiRAC is part of the National e-Infrastructure. AZ and RPN are supported by STFC grant ST/T000341/1 and ST/X000931/1. TB acknowledges funding from the European Union under the European Union's Horizon Europe Research and Innovation Programme 101124282 (EARLYBIRD) and funding by the Deutsche Forschungsgemeinschaft (DFG, German Research Foundation) under grant 325594231, and Germany's Excellence Strategy - EXC-2094 - 390783311. MB has received funding from the European Research Council (ERC) under the European Union's Horizon 2020 research and innovation programme (PROTOPLANETS, grant agreement No. 101002188). Views and opinions expressed are, however, those of the authors only and do not necessarily reflect those of the European Union or the European Research Council. Neither the European Union nor the granting authority can be held responsible for them. All plots in this paper were made with the Python library \texttt{matplotlib} \citep{hunter-2007}. Typesetting was expedited with the use of GitHub Copilot, but without the use of any AI-generated text.

\section*{Data Availability}

Data from our numerical models are available upon reasonable request to the corresponding author.

\bibliographystyle{mnras}
\bibliography{refs}

% \clearpage
% \newpage

\appendix

\section{Two-disk model}
\label{apdx:two-disks}

To set up our two-disk model we first define the cylindrical radius $R=\sqrt{x^2+y^2}$ with respect to a Cartesian coordinate system $\{x,y,z\}$. The midplane is then defined by $z=0$, and the disk is assumed to be vertically isothermal with an aspect ratio $h(R)=H(R)/R$ and a radial surface density profile $\Sigma(R)$, tapered between $R_\text{in}$ and $R_\text{out}$ as

\begin{equation}
	\label{eq:surface-density-taper}
	\Sigma(R) = \Sigma_0\left(\frac{R}{R_\text{in}}\right)^{-1}\frac{1}{1+\exp\left(\frac{R_\text{in}-R}{0.01R_\text{in}}\right)}\frac{1}{1+\exp\left(\frac{R-R_\text{out}}{0.01R_\text{out}}\right)}.
\end{equation}
The gas volume density is then given by \citep[e.g.,][]{nelson-etal-2013}
\begin{equation}
	\label{eq:gas-density}
	\rho_\text{g}(R, z) = \rho_\text{g}^\text{mid}\,e^{-\frac{1}{h^2}\left(1-\frac{R}{\sqrt{R^2+z^2}}\right)},\quad \rho_\text{g}^\text{mid}(R) = \frac{1}{\sqrt{2\pi}}\frac{\Sigma}{H}.
\end{equation}
The disk can then be arbitrarily rotated about the $y$ axis by an angle $\theta_\text{rot}$ with respect to the $xy$ plane by applying the transformation
\begin{equation}
	\label{eq:rotation}
		x' = x \cos\theta_\text{rot} + z \sin\theta_\text{rot},\quad y' = y,\quad z' = -x \sin\theta_\text{rot} + z \cos\theta_\text{rot},
\end{equation}
and finally $R'=\sqrt{x'^2+y'^2}$.

We first define a fixed grid in spherical coordinates $\{r,\theta,\phi\}$, through which we compute $\{x,y,z\}$ at all points. We then rotate the Cartesian coordinate system by $\theta_\text{rot}$ to obtain $\{x',y',z'\}$, define $R'$, and finally compute the gas density $\rho_\text{g}(R', z')$ through the above equations. We set $R_\text{in}=0.2$\,au, $R_\text{out}=4$\,au, and $\theta_\text{rot}=30^\circ$ for the inner disk, and $R_\text{in}=5$\,au, $R_\text{out}=300$\,au, and $\theta_\text{rot}=0$ for the main disk. The total gas density is then given by the sum of the contributions of both disks. Finally, by assuming that the small grains are perfectly coupled to the gas, we write the dust density $\rho_\text{d}^\text{small}$ as
\begin{equation}
	\label{eq:dust-density}
	\rho_\text{d}^\text{small} = \varepsilon (1-X)\rho_\text{g},
\end{equation}
where $X=0.9$ is the coagulation fraction defined in Eq.~\eqref{eq:coagulation-fraction} and $\varepsilon=0.01$ is the dust-to-gas ratio. This approach allows us to model the two-disk system on a single, spherical, regular grid, on which ray-tracing from the central star is straightforward.

\section{Comparison with dynamic radiative transfer}
\label{apdx:radiative-transfer}

To verify how accurate the time-dependent heating/cooling prescription of
this paper is, we carry out a 1D vertical time-dependent radiative transfer
calculation, for a column of gas and dust in approximate vertical hydrostatic
equilibrium orbiting around the star and periodically passing through a
shadow. In this appendix we describe the model, albeit in abbreviated form,
and refer to \citet{dullemond-etal-2002} and \citet{dullemond-etal-2022} for details.

The model is set up on a vertical grid of 90 grid points in the coordinate $z$,
with $z=0$ being the midplane of the disk. The grid spans between $z=0$ and
$z=0.4\,R$, with $R$ being the cylindrical radial distance to the central star
which we set to $R=60\,\mathrm{au}$. For simplicity, we assume mirror symmetry
in the midplane. We are aware of the fact that for disk misalignments unequal to
90 (or 0) degrees, the shadows of the inner disk onto the outer disk will be not
mirror symmetric. However the dynamic models of this paper do not account for
this top/bottom asymmetry, so we keep the 1D model symmetric.

We first compute a steady-state model. The methods used are very similar to
those described by \citet{dullemond-etal-2002}. The only exception is the
convergence algorithm with which the radiative transfer is computed: we use the
moment method with fixed mean opacities only for obtaining a first guess. Then
we use the simple Lambda Iteration method with Ng-acceleration to obtain the
accurate results. The reason why we can afford the Lambda Iteration scheme is
that we focus on the outer disk regions which have much lower optical depth than
the very inner disk regions for which the moment method was developed.

The parameters of our model are $\Mstar=1\,\Msun$, $\Lstar=1\,\Lsun$ with a solar-type spectrum, $\Sigma_{\mathrm{g}}=16.67\,\mathrm{g}/\mathrm{cm}^2$, a dust growth parameter of $X=0.9$, standard DIANA opacities from \optool{} with material density $2.08\,\mathrm{g}/\mathrm{cm}^3$ for small grains with radius $0.1\,\mu\mathrm{m}$ and pebbles with radius $1\,\mathrm{mm}$. We use an incidence irradiation angle of $0.0361$ radians yielding an initial approximate midplane temperature of 18.2\,K, which yields an aspect ratio of $h\simeq 0.066$. The vertical distribution of the gas is then taken to be Gaussian with that pressure scale height. We assume that both the small grains and the pebbles are vertically well-mixed. For the wavelength grid we use 100 gridpoints logarithmically spaced between 0.1\,$\mu$m and 2\,mm. At each wavelength, the radiation field has an angular grid of 20 points in $\mu=\cos\theta$ between $-1\le\mu\le 1$. After the steady-state radiative transfer has converged, the midplane temperature is 16.72\,K for both the gas and the small grains, and 15.9\,K for the pebbles. 

Next we use this steady-state model as the initial condition for a time-dependent computation. As we co-move along the orbit, our 1D column passes through the shadow twice per orbit. The shadow depth is $10^{-6}$, the shadow width is $\Delta\phi=0.48$ radian, corresponding to a pass-through time of 35.5 years. In the time-dependent model, the radiative transfer is assumed to be instant, as the light-crossing time for one scale height is only 2250 seconds (i.e., a factor of half a million times shorter than the crossing time of the gas through the shadow). However, the gas, dust and pebbles are not in thermal equilibrium with the radiation field. The time-dependence of the model is thus the gain/loss of heat from these three components. The physics is described in this paper and in the appendices of \citet{dullemond-etal-2022}. We include the dust--gas coupling, and we assume that the gas, having no continuum opacity, cannot efficiently couple to the radiation field. So the radiative heating and cooling goes solely through the small grains and the pebbles. At each location, the radiative heating rate of the small dust grains (d) or pebbles (p) is
\begin{equation}
  q_{\mathrm{d/p}}^{+}(z) = 4\pi \rho_{\mathrm{d/p}}(z) \int_0^\infty \kappa_{\nu,\mathrm{d/p}}^{\mathrm{abs}}
  J_\nu(z) \text{d}\nu
\end{equation}
in units of energy per time per volume, where $J_\nu(z)$ is the angular-mean
intensity of the radiation field
\begin{equation}
J_\nu(z) = \frac{1}{2}\int_{-1}^{+1} I_{\mu,\nu}(z) \text{d}\mu
\end{equation}
where $I_{\mu,\nu}(z)$ is the radiative intensity as a function of $z$, computed
from the radiative transfer part of the algorithm. The radiative cooling rate is
\begin{equation}
  q_{\mathrm{d/p}}^{-}(z) = 4\pi \rho_{\mathrm{d/p}}(z) \int_0^\infty \kappa_{\nu,\mathrm{d/p}}^{\mathrm{abs}}
  B_\nu(T_{\mathrm{d/p}}(z)) \text{d}\nu
\end{equation}
where $B_\nu$ is the Planck function and $T_{\mathrm{d/p}}(z)$ is the
temperature of the dust or pebble, at a given time. In addition we have the dust--gas (or pebble--gas) thermal coupling $q_{\mathrm{d/p,g}}(z)$ \citep{dullemond-etal-2022}. When $q_{\mathrm{d/p}}^{+}(z)$, $q_{\mathrm{d/p}}^{-}(z)$ and $q_{\mathrm{d/p,g}}(z)$ are not in equilibrium, the temperatures of the three components (dust, pebbles and gas) change, as described in \citet{dullemond-etal-2022}.

\begin{figure}
	\includegraphics[width=\columnwidth]{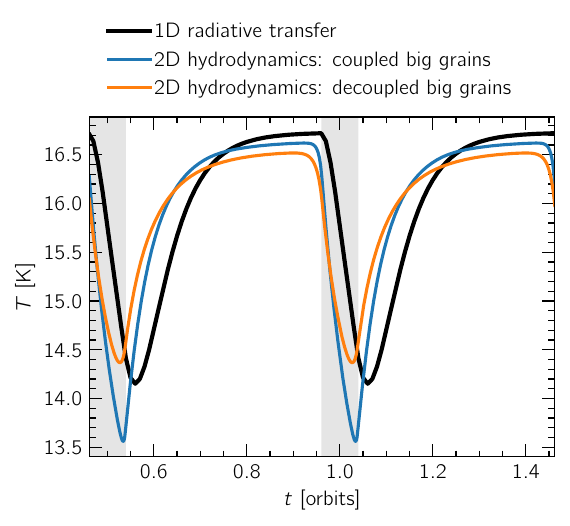}
	\caption{Midplane gas temperature evolution using the dynamic radiative transfer model described in this section (black) and models following the setup in Sect.~\ref{sec:physics-numerics}. The two models agree well in terms of the peak-to-peak temperature variation. The difference in phase owes both to the partial thermal dust--gas decoupling and the lack of radiative diffusion in the 1D model.}
	\label{fig:rt-comparison}
\end{figure}

We can now compare the resulting midplane temperatures as a function of time (or phase) along the orbit, and compare to those computed using the approximate method of this paper. For the latter, we run two dedicated simulations with $\Lstar=1.3\,\Lsun$ in order to match the temperature of 16.72\,K at $R=60$\,au, with the first following the exact setup in Sect.~\ref{sec:physics-numerics} and the second assuming that big grains are thermally decoupled (i.e., $\kappa^\text{big}_\text{R,P}=0$). The result is shown in Fig.~\ref{fig:rt-comparison}, where it can be seen that the two methods match reasonably well. The temperature of the pebbles is always a bit lower than that of the gas and the small grains, but that is to be expected, because the opacity of the big grains is flatter than that of the small grains, and thus tends to lead to lower temperatures.

\section{Azimuthal $T$ variations in \radmc{}}
\label{apdx:azimuthal-temperature}

\begin{figure}
	\includegraphics[width=\columnwidth]{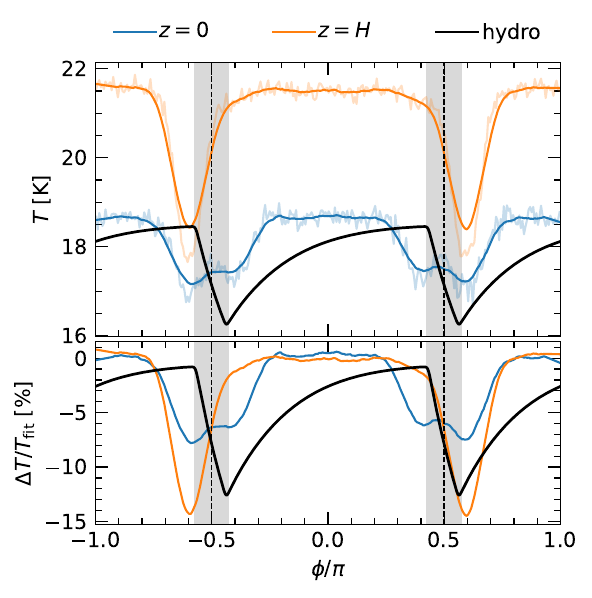}
	\caption{Azimuthal slice at $R=40$\,au of the gas and small dust temperatures, normalized to the azimuthal median for \radmc{} (blue, orange) and to the initial temperature at that radius for \pluto{} (black). The latter shows a wider spread of temperatures due to the azimuthal advection of thermal energy. \radmc{} and \pluto{} show excellent agreement in background temperature at the midplane.}
	\label{fig:shadow-shift}
\end{figure}

As mentioned in Sect.~\ref{sub:radmc-setup}, \radmc{} is a hydrostatic code and therefore does not capture the azimuthal advection of thermal energy. This results in artificially narrower temperature fluctuations compared to our hydrodynamical simulations. To highlight this effect we show the azimuthal profiles of $T_\text{small}\approx T_\text{gas}$ from the \radmc{} output of our fiducial model at $t=1000\,P_0$ and $R=40$\,au in Fig.~\ref{fig:shadow-shift} and compare it to the \pluto{} output at the same timestamp and radius.

As seen on that figure, deviations from said background are found only around the shadowed region in \radmc{} whereas they span the full azimuthal range in \pluto{}. The \radmc{} model also shows weaker deviations compared to the background near the midplane compared to \pluto{}. Nevertheless, the two codes agree very well at the midplane in terms of the background temperature.

\bsp	% typesetting comment
\label{lastpage}
\end{document}